\documentclass[a4paper,onecolumn,11pt,accepted=2023-03-05]{quantumarticle}

\pdfoutput=1
\usepackage[utf8]{inputenc}
\usepackage[english]{babel}
\usepackage[T1]{fontenc}
\usepackage{braket}
\usepackage{amsmath}
\usepackage{amssymb}
\usepackage{amsthm}
\usepackage{hyperref}
\usepackage{subcaption}
\usepackage{placeins}
\newtheorem{definition}{Definition}
\DeclareMathOperator{\Tr}{Tr}

\usepackage{tikz}
\usepackage{lipsum}

\newtheorem{theorem}{Theorem}

\begin{document}

\title{Entanglement Trajectory and its Boundary}

\author{Ruge Lin}
\affiliation{Quantum Research Centre, Technology Innovation Institute, United Arab Emirates.}
\affiliation{Departament de F\'isica Qu\`antica i Astrof\'isica and Institut de Ci\`encies del Cosmos, Universitat de Barcelona, Spain.}
\orcid{0000-0002-9297-7622}

\begin{abstract}

In this article, we present a novel approach to investigating entanglement in the context of quantum computing. Our methodology involves analyzing reduced density matrices at different stages of a quantum algorithm's execution and representing the dominant eigenvalue and von Neumann entropy on a graph, creating an "entanglement trajectory." To establish the trajectory's boundaries, we employ random matrix theory. Through the examination of examples such as quantum adiabatic computation, the Grover algorithm, and the Shor algorithm, we demonstrate that the entanglement trajectory remains within the established boundaries, exhibiting unique characteristics for each example. Moreover, we show that these boundaries and features can be extended to trajectories defined by alternative entropy measures. The entanglement trajectory serves as an invariant property of a quantum system, maintaining consistency across varying situations and definitions of entanglement. Numerical simulations accompanying this research are available via open access.

\end{abstract}

\maketitle

\section{Introduction}

An appropriate dose of entanglement can be seen as the resource of the quantum advantage for computation \cite{jozsa2003role,orus2004universality}. It can not be too low since quantum states whose von Neumann entropy only scales logarithmically with the number of qubits can be described in terms of Matrix Product States \cite{vidal2003efficient}. This means the amount of entropy in the system along the quantum algorithm should be large; otherwise, an efficient algorithm would exist to solve the same problem using Tensor Networks. It can not be too high either; otherwise, quantum resources can be replaced with a fair coin since maximum entanglement usually implies maximum randomness \cite{gross2009most}. 

Therefore, all quantum states that present a potential for computational speedup lie in the separation between "not entangled enough" and "too entangled," which is very narrow since almost all quantum states are extremely entangled \cite{gross2009most}. This article introduces a new approach to studying this gap, referred to as the entanglement trajectory. Since genuine multipartite entanglement measures are not available for states with a large number of qubits \cite{bengtsson2017geometry}, for simplicity, this article only focuses on bipartite pure quantum states. The result is already fruitful under these constraints. For such a quantum system, the value of the von Neumann entropy originates from the eigenvalues $\lambda_i$ of the reduced density matrix $\rho_A$, where the subsystem $A$ is partially traced over its complementary system. In our experience, these eigenvalues usually consist of an isolated dominant member $\lambda_0$, while the rest are numerically approached to $0$. This particular behaviour provides the initial motivation for this paper. We ask, what happens when we plot the entanglement trajectory for a quantum process performing a computation, with the dominant eigenvalue on the $x$-axis and the von Neumann entropy on the $y$-axis? How does the trajectory change when using other measures of entanglement on the $y$-axis? What additional information can be gained about the system?

\section{Examples from quantum computation}\label{examples}

We briefly summarize what we expect to see before studying different quantum systems. We use the von Neumann entropy defined on the natural logarithm $E=-\sum \lambda_i\ln \lambda_i$ in this article, with $\lambda_i \geq 0$ and $\sum \lambda_i =1$. When the size of the subsystem $A$ is $\alpha$ and the size of its complementary system is $\beta\geq\alpha$, for the dominant eigenvalue, we should have $1\geq\lambda_0\geq\frac{1}{\alpha}$ and the maximum entropy is $\ln{\alpha}$.

Also, when only two non-zero eigenvalues exist, the entropy reaches its lower bound

\begin{equation}
f_1=-\lambda_0\ln\lambda_0-\left(1-\lambda_0\right)\ln\left(1-\lambda_0\right).
\end{equation}

This $f_1$ will be plotted in a gray dash-dotted line throughout this article.

When $\lambda_0=\frac{1}{w}$ and $w$ is an integer with $\alpha \geq w \geq 2$, the entropy reach its lower bound $f_2$ when there is $w$ non-zero eigenvalues and all of them are $\frac{1}{w}$, with

\begin{equation}
f_2=-\ln{\lambda_0},
\end{equation}
which will be plotted in a gray dashed line throughout this article. Notice that $f_1$ and $f_2$ do not depend on the size of the subsystem $\alpha$ and will be plotted in every figure in this article when the $y$-axis is the von Neumann entropy.

Furthermore, the entropy will reach its upper bound $f_3$ when the rest of the eigenvalues are all equal and sum up to $1-\lambda_0$, when $\alpha\sim\alpha-1$, we have

\begin{equation}
f_3=\left(1-\lambda_0\right)\ln{\alpha}+f_1,
\end{equation}
which will be plotted in a gray solid line in every figure with one $\alpha$ and with the von Neumann entropy as the $y$-axis through this article.

\begin{figure}
\centering
\includegraphics[width=0.48\linewidth]{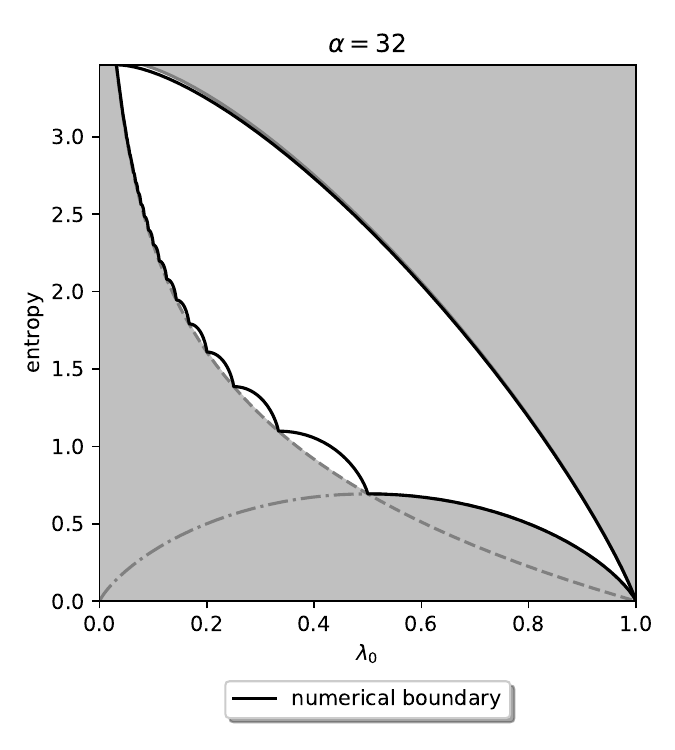}
\caption{The numerical boundary versus the analytical boundary constrained by $f_1$, $f_2$ and $f_3$. The numerical boundary is calculated as follow: take $\lambda_0=0.4$, its lower bound is given by $-0.4\ln\left(0.4\right)-0.4\ln\left(0.4\right)-0.4\ln\left(0.2\right)$ and its upper bound is given by $-0.4\ln\left(0.4\right)-31\times\frac{1-0.4}{31}\ln\left(\frac{1-0.4}{31}\right)$.}
\label{bounds}
\end{figure}

The comparison between the boundary formed by $f_1$, $f_2$ and $f_3$ and the actual numerical result is shown in FIG. \ref{bounds}. As we can see, the numerical boundary can be approximated with these three analytical lines while staying strictly within the area confined by them. At this point, the boundary is a direct result of the definition of von Neumann entropy. This boundary is tight and does not rely on the purity of quantum states. It is impossible to provide a counter-example.

In this section, we present three quantum algorithms and track their entanglement trajectories. They are all well-studied and with explicit numerical implementation on the open-access platform Qibo \cite{qibo_paper,qibojit_paper}. Using them as examples is both convincing and convenient. The code which generates every figure in this article is on GitHub \cite{Github}. We must highlight that in each figure, the lines that connected data points do not have any physical meaning and serve only the purpose of visualizing the sequential order between quantum states. The transitions between quantum states are not continuous and can never be presented as a path connecting two data points. Otherwise, it will be classical. For the result, we find out these trajectories present a unique feature for each quantum algorithm and can be used as "fingerprints." However, since the eigenvalues of these density matrices do not follow any analytical distribution, it will be difficult to guarantee a further mathematical explanation.

\subsection{Adiabatic quantum computation}\label{adiabtic}

In Ref. \cite{orus2004universality}, the authors have studied the entanglement properties of the adiabatic quantum algorithm for solving the NP-complete Exact Cover (EC) problem, which is a particular case of $3$-SAT problem. An Exact Cover instance is characterized by a set of clauses containing $3$ bits that are considered satisfied if one is in position $1$ while the other remains at $0$. The solution of this instance is one bit-string that fulfills all the clauses at the same time. To explain the adiabatic quantum algorithm in a nutshell \cite{albash2018adiabatic}, the quantum system is initialized with an "easy" Hamiltonian $\mathcal{H}_0$, which is a magnetic field in the $x$ direction, where the ground state is an equal superposition of all possible computational states. The system is then adiabatically evolved to a "complicated" Hamiltonian $\mathcal{H}_p$ that encodes an instance of EC problem, where the ground state is the solution. During the adiabatic evolution, the system can be described using a Hamiltonian $\mathcal{H}_s$, an interpolation between $\mathcal{H}_0$ and $\mathcal{H}_p$,

\begin{equation}
\mathcal{H}_s=\left(1-s\right)\mathcal{H}_0+s\mathcal{H}_p\hspace{0.5cm}\text{with}\hspace{0.5cm}s\in[0,1].
\end{equation}

\begin{figure}
\centering
\includegraphics[width=0.48\linewidth]{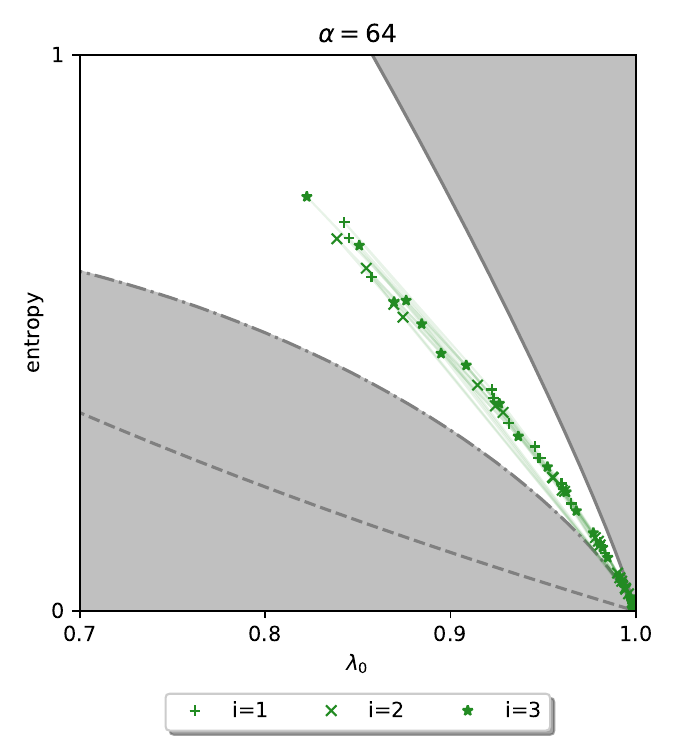}
\caption{The entanglement trajectories for solving three instances of EC problem with $n=12$, we took three random bi-partition (where the subsystem $A$ is a random half of qubits) of each instance. Different instances are plotted with markers $+$, $\times$ and $\star$.}
\label{EC_bis}
\end{figure}

In the original paper, the authors tracked the von Neumann entropy of different equal bi-partition (using half of the qubits as the subsystem and $\alpha=2^{n/2}$) of the ground state of $\mathcal{H}_s$ during the adiabatic evolution for solving EC problem. They found out that the entanglement always reaches its pic at $s\sim 0.7$ while the pic scales $\mathcal{O}\left(n\right)$ for different sizes of systems. Their results imply that this pic corresponds to an undergo phase transition; therefore, a classical computer cannot efficiently simulate the adiabatic evolution. However, the difficulty of simulation does not signify a quantum advantage. Though it is almost impossible to prove, as pointed out by Ref. \cite{dickson2011does}, the gap between the ground energy and the first excited energy of $\mathcal{H}_s$ is believed to be exponentially small for an NP-complete problem \cite{vznidarivc2006exponential}. Evaluation requires an exponentially long time to reach the correct solution; even an optimal adiabatic scheduling function for $s$ can only produce a quadratic speed-up \cite{albash2018adiabatic}. The numerical simulation has been re-implemented in Qibo, and the code is on GitHub \cite{Github_EC}. In this section, we revisit their study and track the entanglement trajectory over random bi-partition of the ground state with the interval of $s$ as $0.1$. The results are shown in FIG. \ref{EC_bis}. As we can see, the data points collected are distinct for different instances and bi-partition, but they almost stay on the same path.

\subsection{Grover algorithm}

The quantum search algorithm for an unordered database was first introduced by Grover \cite{grover1996fast}. The search space is initialized as the equal superposition of all possible solutions (usually the equal superposition of all computation basis). A Grover step consists of two parts: the oracle that encodes the description of the problem and inverts the sign of the solutions, followed by the diffusion operator that amplifies the probability of measuring the solutions. The repetition of Grover steps brings the probability of finding the right solutions close to $1$. Then in Ref. \cite{orus2004universality}, the authors have proved that the entropy in the register along the Grover algorithm is bounded by $1$ after each Grover step. Consequently, the quantum advantage of the Grover algorithm should be expected from entanglement present in the middle of the oracle. In this section, we compare the entanglement trajectory of two search algorithms. The bi-partition splits between the search space (where the diffusion operator takes place) and its complementary space.

\begin{figure}
\centering
\includegraphics[scale=0.9]{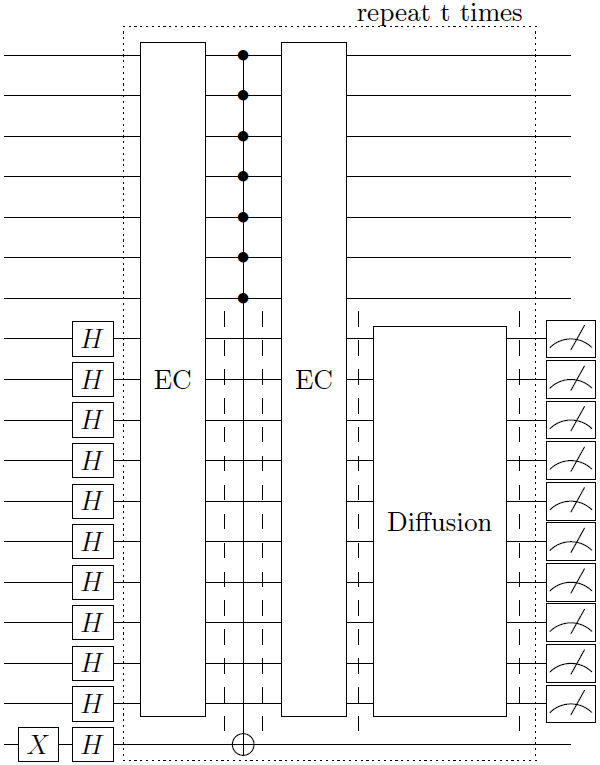}
\caption{The quantum circuit for Grover algorithm to solve one instance of EC problem with $n=10$ and $7$ clauses. The number of bits and clauses are chosen such that two Grover examples share the same number of qubits and can be traced over subsystems of the same size. We sample $\rho_A$ in each iteration before and after the central multi-Toffoli gate and the diffusion operator as indicated with dashed lines. In this circuit, we have $t=25$.}
\label{Grover_circuit_3sat}
\end{figure}

The first example is to use the Grover algorithm to solve the EC problem mentioned in the section \ref{adiabtic}. The oracle has the same number of ancillas as clauses to ensure that all clauses are satisfied. For each clause, a CNOT gate from each qubit to its ancilla, followed by a multi-Toffoli gate controlled by the three qubits targeting the ancilla, activates the ancilla only if the clause is satisfied. The circuit is shown in FIG. \ref{Grover_circuit_3sat}, and the implementation in Qibo is provided on GitHub \cite{Grover_EC}. Due to its high entanglement, this quantum circuit can not be efficiently simulated on classical hardware. But it does not imply a quantum advantage. The quantum algorithm scales $\mathcal{O}\left(2^{0.5n}\right)$ \cite{dalzell2022mind} while the best classical algorithm scales $\mathcal{O}\left(2^{0.39n}\right)$ \cite{hansen2019faster}.

\begin{figure}
\centering
\includegraphics[scale=0.9]{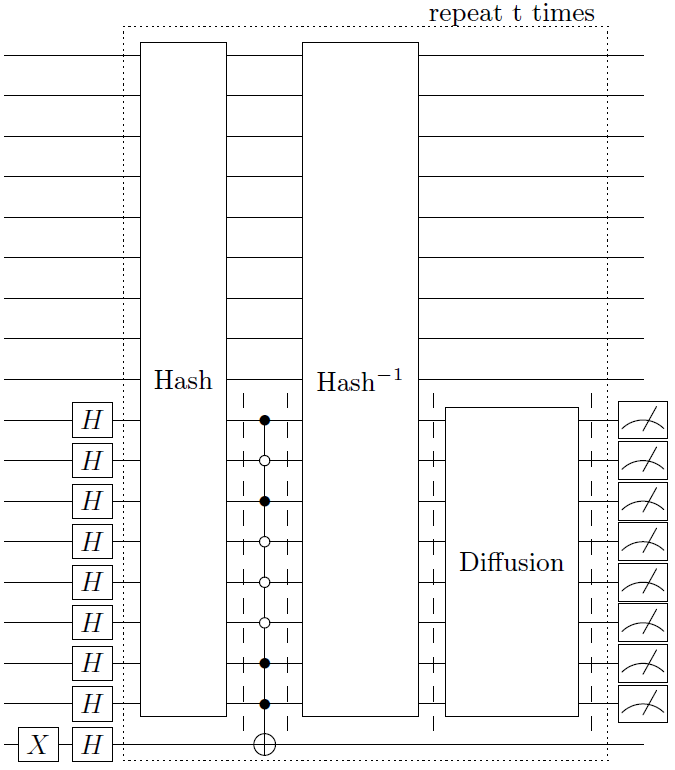}
\caption{The quantum circuit of Grover algorithm to find the pre-image of a Hash function with a given cipher-text. The oracle consists of a unitary that encodes the Hash function, the inverse of this unitary, with a multi-Toffoli gate in the middle that encodes the cipher-text, which is $10100011$ in this figure. We sample $\rho_A$ in each iteration before and after the multi-Toffoli gate and the diffusion operator as indicated with dashed lines. This cipher-text has $2$ pre-images, and the number of Grover iterations needed is $t=8$.}
\label{Grover_circuit_hash}
\end{figure}

The second example \cite{ramos2021quantum} is the application of the Grover algorithm to find the pre-images of a toy Sponge Hash function \cite{bernstein2008chacha}. Their circuit consists of $18$ qubits, where $n=8$ are initialized as a uniform superposition of all states, referred to as the search space of Grover's algorithm. Like the previous example, this Grover circuit can not be efficiently simulated with a classical computer due to the high entanglement in the middle of the oracle, as indicated in their paper. However, the quantum algorithm does provide an advantage. For a given function, if there exists a classical algorithm that can output the pre-images of an arbitrary cipher-text more efficiently than blind search, it can not be qualified as a Hash function. And Grover algorithm performs blind search with a quadratic speed-up. Details are provided in their paper and their GitHub page \cite{Github_hash}. In this section, besides the entanglement, we have also tracked the entanglement trajectory. A simplified circuit is shown in FIG. \ref{Grover_circuit_hash}. 

\begin{figure}
\centering
\includegraphics[width=0.48\linewidth]{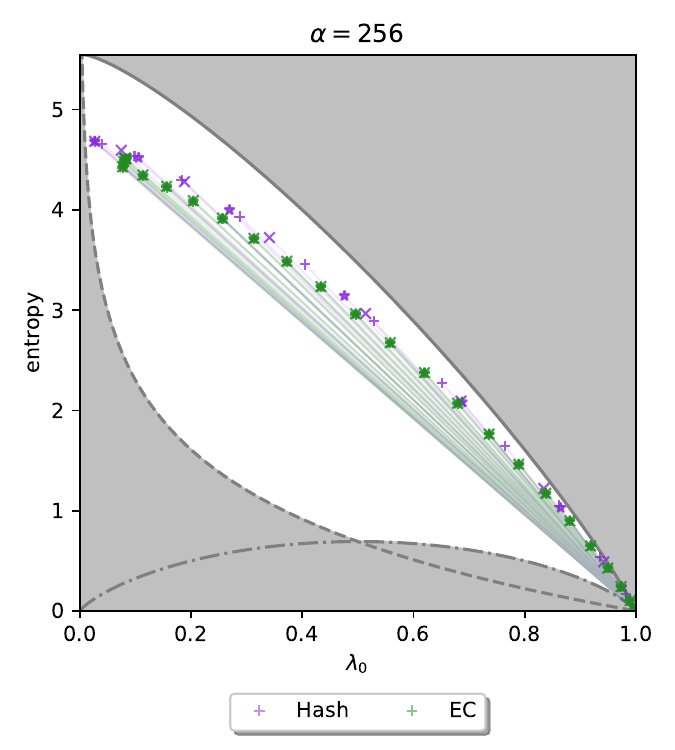}
\caption{The entanglement trajectory of Grover circuits to solve three instances (with $n=10$ and $7$ clauses) of EC problem (plotted in green markers $+$, $\times$ and $\star$) and three cipher-texts (with $1$, $2$ and $3$ pre-images) of the same Hash function (plotted in violet markers $+$, $\times$ and $\star$).}
\label{Grover_bis}
\end{figure}

The comparison of the entanglement trajectory of these two Grover algorithms can be found in FIG. \ref{Grover_bis}. The sizes of subsystem $\alpha=2^8$ and the complementary system $\beta=2^{10}$ are the same. As we can see, although the density of data points is distinct, two Grover algorithms that solve different problems share a similar path. In particular, the trajectory of the Grover algorithm that solves the Hash function is slightly longer and slightly above the one that solves the EC problem.

\subsection{Semi-classical Shor algorithm}

The original paper by Shor \cite{shor1999polynomial} stated that the factorization problem could be reduced to order finding. Then, Shor proposed a quantum order-finding algorithm that runs in polynomial time. To factorize a number $N=p\times q$, a number $1<a<N$ is randomly chosen and the quantum algorithm finds a number $r$ such that $a^r=1\mod N$. If $r$ is neither odd nor verifying $a^{r/2} = -1 \mod N$, $p$ and $q$ can be recovered from $r$.

Since the efficiency of the Shor algorithm is due to the efficiency of the quantum Fourier transform (QFT), previous research mainly focused on studying the entanglement entropy before the application of the QFT and in between the register of the QFT and the quantum modular exponentiation (QME) \cite{orus2004universality,kendon2004entanglement}. It is concluded that with this partition, the reduced density matrix is diagonal with $m$ non-zero elements, which all equals $\frac{1}{m}$ with positive integer $m$. This behaviour matches the scenario that $f_2$ is deduced; therefore, its entanglement trajectory (one data point for a given $N$ and $a$) should locate on $f_2$. And we verify this property with numerical simulation using Qibo, which is available on GitHub \cite{Github_shor}. They also showed that $m$ is $\mathcal{O}\left(r\right)$, which indicates that the entropy scales $\mathcal{O}\left(\ln{r}\right)$.

\begin{figure}
\centering
\includegraphics[scale=0.9]{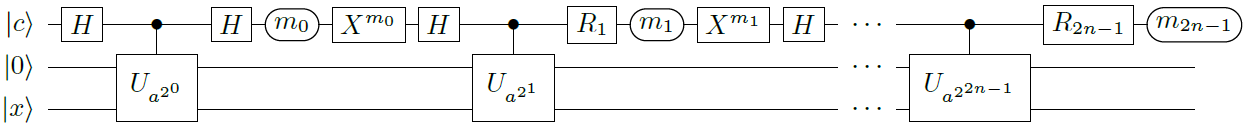}
\caption{The semi-classical circuit for the Shor algorithm using $2n+3$ qubits. Details are provided in Ref. \cite{beauregard2002circuit}. $\ket{c}$ is the controlling qubit that performs the QFT. Register $\ket{x}$ consists of $n$ qubits and encodes the output of the QME, and is initialized as $\ket{1}$ (it means the least significant qubit is $\ket{1}$ while others in $\ket{0}$). Register $\ket{0}$ consists of $n+2$ qubits and serves as ancillas.}
\label{Shor_circuit}
\end{figure}

\begin{figure}
\centering
\includegraphics[scale=0.9]{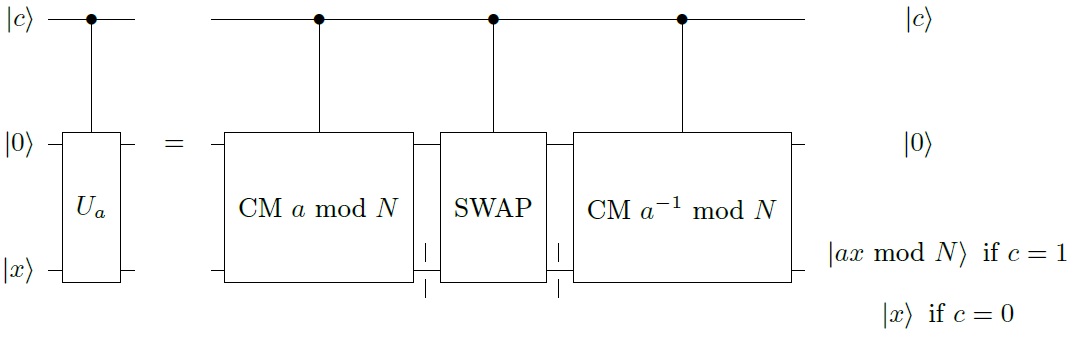}
\caption{The decomposition of the controlled QME. CM is the circuit for the controlled multiplication. Details are provided in Ref. \cite{beauregard2002circuit}. The reduced density matrices $\rho_A$ are sampled before and after each time the controlled SWAP operator in the middle is applied, and the subsystem $A$ is $\ket{x}$, as shown with dashed lines.}
\label{Shor_circuit_ME}
\end{figure}

\begin{figure}
\centering
\includegraphics[width=0.48\linewidth]{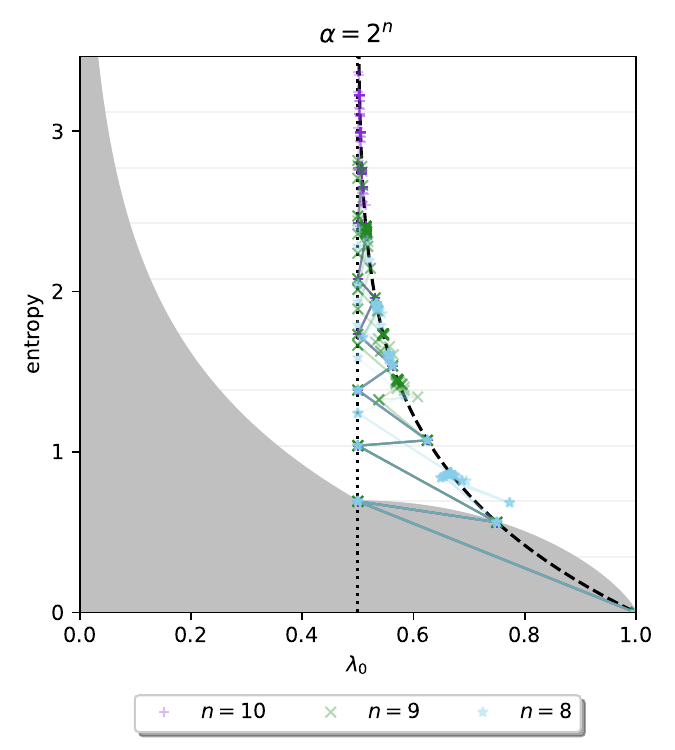}
\caption{The entanglement trajectories of different semi-classical Shor circuits. For each $n$, three different $N$ and three different $a$ were chosen (nine combinations).  Data that were taken after the controlled SWAP gate cluster on $\lambda_0=\frac{1}{2}$ (plotted as a black vertical dotted line) with an interval of $\frac{1}{2}\ln{2}$ (plotted with gray horizontal lines). And data that were taken before the controlled SWAP gate accumulate on the curve $f_{shor}$ (plotted as a black dashed line). Data points are plotted with high transparency to show the concentration of the trajectory.}
\label{Shor_bis}
\end{figure}

In this section, we investigate the entanglement trajectory of the Shor algorithm with the most up-to-date version for minimizing the number of qubits provided by Beauregard \cite{griffiths1996semiclassical,parker2002entanglement,beauregard2002circuit}, which only needs $2n+3$ qubits. The QFT is performed semi-classically with one controlling qubit, as shown in FIG. \ref{Shor_circuit}. The numerical simulation with Qibo is also available on GitHub \cite{Github_shor}. Since the space of QFT has been reduced to one qubit, the entanglement between this qubit and the rest of the system is upper bounded by $1$, and it would be more convincing to study this circuit with another partition. We use the bi-partition between the register that stores the output of the QME ($n$ qubits) and the rest of the qubits to track the entanglement trajectory and sample $\rho_A$ in the middle of each QME, as shown in FIG. \ref{Shor_circuit_ME}. The size of the subsystem is $\alpha=2^{n}$, and the size of the complementary system is $\beta=2^{n+3}$. The result is shown in FIG. \ref{Shor_bis}. The entanglement trajectories concentrate on a particular path connecting $\lambda_0=\frac{1}{2}$ with an interval of $\frac{1}{2}\ln{2}$ and

\begin{equation}
f_{shor}=-x\ln{x}-\left(1-x\right)\ln\left(x-\frac{1}{2}\right),
\end{equation}
despite the difference between system sizes and inputs. The maximum entanglement is located at $\lambda_0=\frac{1}{2}$. We do not have an explicit interpretation of the $\frac{1}{2}\ln{2}$ interval nor $f_{shor}$. As we can see from the numerical output, some data points are located quite far from this approximation. The interval is derived from the fact that the eigenvalues of $\rho_A$ after the controlled SWAP operator frequently consist of $\lambda_0=\frac{1}{2}$ and $2^m$ non-zero elements of $\frac{1}{2^m}$, with positive integer $m$. And the formula $f_{shor}$ is derived from the fact that the eigenvalues of $\rho_A$ before the controlled SWAP operator frequently consist of $\lambda_0=\frac{m+1}{2m}$ and $m-1$ non-zero elements of $\frac{1}{2m}$, with positive integer $m$.

\section{Mathematics Background}

In this section, we briefly introduce two key ingredients for the asymptotic analysis of entanglement trajectory. As most quantum states are random, it is natural that the entanglement of a physical system should concentrate on the entanglement obtained with random density matrices, which can be generated by partially tracing a larger random state \cite{braunstein1996geometry,sommers2004statistical}. It is proven that the law of a random density matrix is the law of a Wishart matrix conditioned by trace $1$ \cite{nechita2007asymptotics,majumdar2010extreme}. Therefore, the first ingredient we present is one theorem in random matrix theory defined on the Wishart ensemble, the Marchenko Pastur distribution. 

The second ingredient consists of analyzing the isolated dominant eigenvalue when the coefficients in the reduced density matrix do not centre at $0$ using a basic theorem in linear algebra. This method mainly focuses on the expectation of the dominant eigenvalues rather than their distribution. It is sensitive to the condition that we generate random reduced density matrices (by setting the mean and the variance of the random variables). Also, it has a more straightforward analytical expression that helps us derive the relation between the entropy and the dominant eigenvalue.

\subsection{Marchenko Pastur distribution}\label{MPD}

We introduce the Marchenko Pastur distribution \cite{nechita2007asymptotics,feier2012methods,livan2018introduction,bai1999kmethodologies,haagerup2003random,potters2020first}, which describes the asymptotic behaviour of eigenvalues of a Wishart matrix. The theorem is named after mathematicians Vladimir Marchenko and Leonid Pastur, who proved it in 1967 \cite{marvcenko1967distribution}.

\begin{definition}{Wishart matrix}

Let $X$ be an $\alpha \times \beta$ complex matrix whose elements are complex normal $\mathcal{N}_{\mathbb{C}}\left(0,1\right)$ random variables that are independently identically distributed (i.i.d). The $\alpha\times\alpha$ matrix $XX^{\dagger}$ is referred to as a Wishart randomly generated matrix with parameters $\alpha$ and $\beta$. All Wishart matrices form the Wishart ensemble \cite{wishart1928generalised}.

\end{definition}

\begin{definition}{Empirical spectral distribution, (ESD)}

Given an $\alpha\times\alpha$ symmetric or Hermitian matrix $M_{\alpha}$, consider its $\alpha$ real eigenvalues $\lambda_0, ..., \lambda_{\alpha-1}$. To study their distribution, we form the empirical spectral distribution,

\begin{equation}
\mu_{\alpha}=\frac{1}{\alpha}\sum_{i=0}^{\alpha-1}\delta_{\lambda_i},
\end{equation}
with $\delta_{\lambda_i}\left(x\right)$ being the indicator function $\mathbb{1}_{\lambda_i\leq x}$ \cite{anderson2010introduction}.

\end{definition}

\begin{theorem}{Marchenko Pastur distribution, (MPD)}

If $X$ denotes an $\alpha \times \beta$ random matrix whose entries are i.i.d $\mathcal{N}_{\mathbb{C}}\left(0,\sigma\right)$ random variables. Let $Y_{\beta}=\frac{1}{\beta}XX^{\dagger}$. It is a Wishart matrix. And let $\lambda_0, ..., \lambda_{\alpha-1}$ be eigenvalues of $Y_{\beta}$. Assume that $\alpha,\beta\rightarrow\infty$ and the ratio $\alpha/\beta\rightarrow\lambda\in\left(0,+\infty\right)$.

Then the ESD of $Y_{\beta}$, $\mu_{\alpha}$ converges weakly, in probability, to the distribution with density function $\mu$ given by

\begin{equation}
\mu =
\begin{cases}
\left(1-\frac{1}{\lambda}\right)\mathbb{1}_{0}+\nu & \text{if \hspace{0.3cm} $\lambda > 1$} \\
\nu & \text{if \hspace{0.3cm} $0 < \lambda \leq 1$}\\
\end{cases}\,,
\label{eq_mu_MPD}
\end{equation}
and

\begin{equation}
d\nu(x)=\frac{1}{2\pi\sigma^2} \frac{\sqrt{\left(\lambda_{+} -x\right)\left(x-\lambda_{-}\right)}}{\lambda x} \mathbb{1}_{x\in[\lambda_-,\lambda_+]}dx\,,
\label{eq_nu_MPD}
\end{equation}
with

\begin{equation}
\lambda_{\pm}=\sigma^2\left(1\pm \sqrt{\lambda}\right)^2\,.
\label{eq_lambda_MPD}
\end{equation}
\end{theorem}

\begin{figure}
\centering
\includegraphics[width=0.60\linewidth]{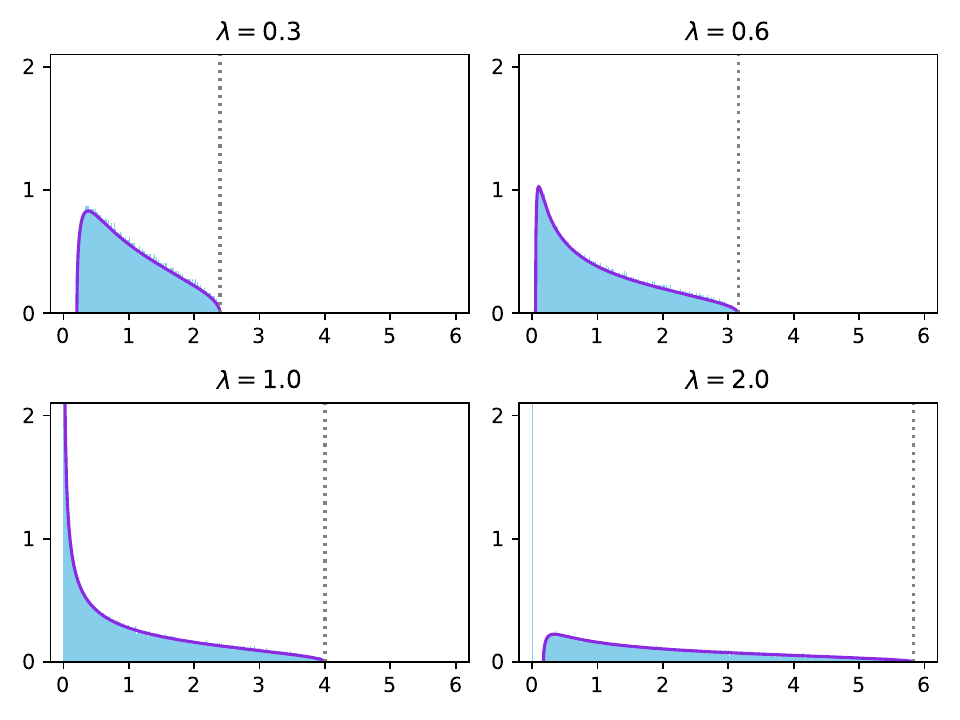}
\caption{The MPD (in violet) versus the numerical results (in blue) for different $\lambda$ with $\alpha=10000$. The gray dotted line indicates $\lambda_+$.}
\label{fig_MPD}
\end{figure}

Ref. \cite{feier2012methods} contains the proof for the MPD using the moment approach. In short, the MPD predicts the distribution of the eigenvalues of a Wishart matrix by giving the analytical outline of its normalized histogram. Let $X$ be an $\alpha\times \beta$ random matrix with variables $\mathcal{N}(0,1)$, $\lambda=\alpha/\beta$. Then the normalized histogram of the eigenvalues of $Y_{\beta}=\frac{1}{\beta}XX^{\dagger}$ is approximated by $\left(x,\frac{1}{2\pi} \frac{\sqrt{\left(\lambda_{+} -x\right)\left(x-\lambda_{-}\right)}}{\lambda x}\right)$ with $\lambda_{\pm}=\left(1\pm \sqrt{\lambda}\right)^2$ as shown in FIG. \ref{fig_MPD}. In this article, the $\lambda_+$ will be indicated with a gray dotted line (can be horizontal or vertical) in the figure when specific $\alpha$ and $\beta$ are given.

\subsection{Dominant eigenvalue of decentralized Wishart matrices}\label{l0}

If $X$ denotes an $\alpha \times \beta$ complex matrix whose entries are i.i.d random variables with mean $\gamma\neq 0$ and variance $\sigma^2$, let $Y_{\beta}=\frac{1}{\beta}XX^{\dagger}$. We call $Y_\beta$ a decentralized Wishart matrix. And let $\lambda_0, ..., \lambda_{\alpha-1}$ eigenvalues of $Y_{\beta}$, with $\lambda_0 \geq ... \geq \lambda_{\alpha-1}$.

We can rewrite $X$ into

\begin{equation}
X=H+\widetilde{X},
\end{equation}
and

\begin{equation}
X^\dagger=H^\dagger+\widetilde{X}^\dagger,
\end{equation}
where $H$ is an $\alpha\times\beta$ with all elements $\gamma$, and $\widetilde{X}$ an $\alpha\times\beta$ i.i.d random matrix with mean $0$ and variance $\sigma^2$.

And we also have the relation

\begin{equation}
H=X+\left(-\widetilde{X}\right),
\end{equation}
and

\begin{equation}
H^\dagger=X^\dagger+\left(-\widetilde{X}^\dagger\right),
\end{equation}
since the elements of $\widetilde{X}$ are centred at $0$.

We define the spectral norm \cite{meyer2000matrix, belitskii2013matrix} of a matrix $P$ as

\begin{equation}
||P||=\sqrt{\lambda_{max}\left(P^\dagger P\right)},
\end{equation}
which satisfies the inequality $||P+P'||\leq ||P||+||P'||$.

Therefore, we have

\begin{equation}
||X^\dagger||\leq ||H^\dagger|| + ||\widetilde{X}^\dagger||,
\end{equation}
which means

\begin{equation}
\sqrt{\lambda_{max}\left(XX^\dagger\right)}\leq\sqrt{\lambda_{max}\left(HH^\dagger\right)}+\sqrt{\lambda_{max}\left(\widetilde{X}\widetilde{X}^\dagger\right)}.
\end{equation}

In the same way,

\begin{equation}
||H^\dagger||\leq||X^\dagger||+||-\widetilde{X}^\dagger||,
\end{equation}
implies

\begin{equation}
\sqrt{\lambda_{max}\left(HH^\dagger\right)}\leq\sqrt{\lambda_{max}\left(XX^\dagger\right)}+\sqrt{\lambda_{max}\left(\widetilde{X}\widetilde{X}^\dagger\right)}.
\end{equation}

So we have when $\lambda_{max}\left(HH^\dagger\right)\gg\lambda_{max}\left(\widetilde{X}\widetilde{X}^\dagger\right)=\lambda_+$, $\lambda_{max}\left(XX^\dagger\right)\sim\lambda_{max}\left(HH^\dagger\right)$.

Then $HH^\dagger$ is an $\alpha\times\alpha$ matrix with all elements equal to $\beta|\gamma|^2$, which means all its eigenvalues are $\alpha\beta|\gamma|^2$.

\begin{figure}
\centering
\includegraphics[width=0.48\linewidth]{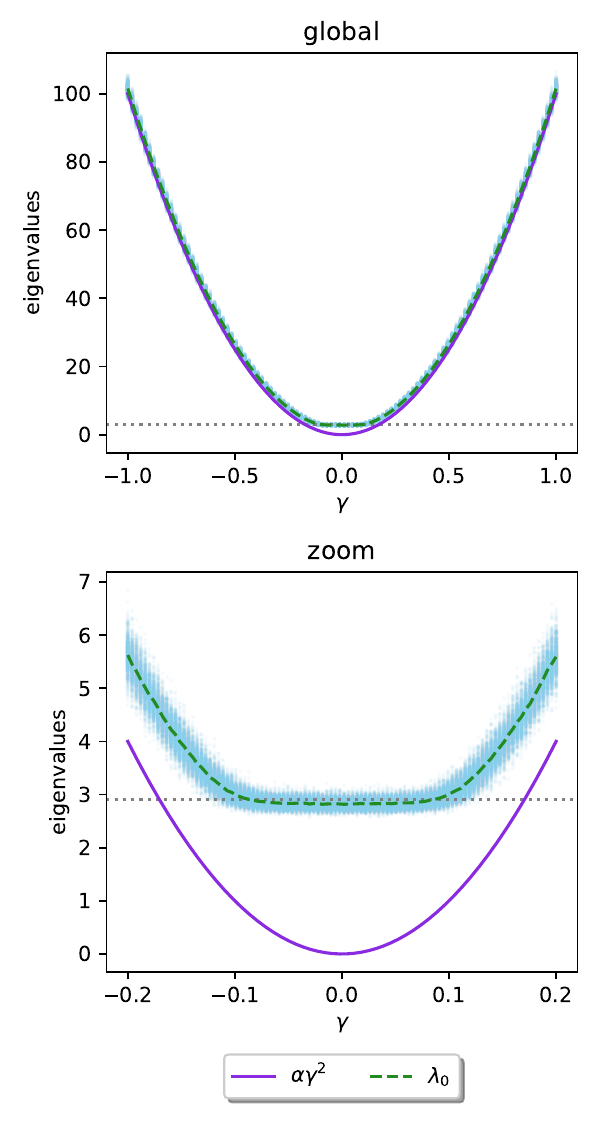}
\caption{The convergence of $\lambda_0$ towards $\alpha\gamma^2$ as $\alpha=100$, $\beta=200$ and $\gamma$ varies from $0$ to $1$. Each $\lambda_0$ is the mean (raw data are plotted in blue) of the dominant eigenvalues over $500$ random matrices $Y_\beta$ generated by the entries of $X$ following the normal distribution $\mathcal{N}\left(\gamma,1\right)$. We have $\lambda=\frac{1}{2}$, and $\lambda_+=\left(1+\frac{\sqrt{2}}{2}\right)^2$ is plotted in gray dotted line. The minor difference lies between $\lambda_0$ and $\lambda_+$ when $\gamma\sim 0$ is due to the effect of the Tracy-Widom distribution \cite{tracy1996orthogonal,tracy2002distribution,johnstone2001distribution,chiani2014distribution} and can be generalized with a phase transition \cite{baik2005phase,vznidarivc2012subsystem,pandey2010correlated,vinayak2014spectral}. Since this phenomenon is highly non-analytical also numerically negligible for our study, we will not elaborate in detail.}
\label{mean_MPD}
\end{figure}

Finally we have $\lambda_0=\lambda_{max}\left(Y_\beta\right)=\lambda_{max}\left(\frac{1}{\beta}XX^\dagger\right)=\alpha|\gamma|^2$, as shown in FIG. \ref{mean_MPD}, while the remaining eigenvalues still follow the MPD \cite{bouchaud2015financial}. We keep the notation $\gamma^2$ instead of $|\gamma|^2$ for simplicity.

\section{Asymptotic Analysis}

In this section, we first demonstrate how to use the section \ref{MPD} to recover a famous theorem \cite{page1993average}. After that, we introduce the element from the section \ref{l0} to give a new asymptotic result for the entanglement trajectory. Then we compare our analysis with previous examples in the section \ref{examples}.

\subsection{Average entropy of a subsystem}

Here we consider random density matrices obtained by partially tracing larger random pure states. There is a strong connection between these random density matrices and the Wishart ensemble of random matrix theory, that can be used to determine the formula of entropy by Don Page in 1993 \cite{nechita2007asymptotics,page1993average,bengtsson2017geometry,sen1996average, pal2020probing, zyczkowski2001induced,hayden2006aspects}.

\begin{theorem}{Average entropy of a subsystem}

If a quantum system of Hilbert space dimension $\alpha\beta$ is in a random pure state, the average von Neumann entropy $E$ of a subsystem of dimension $\alpha \leq \beta$ is $\ln{\alpha}-\frac{\alpha}{2\beta}$ for $1 \leq \alpha \leq \beta$.

\end{theorem}

We can use the MPD to reproof it.

\begin{proof}

Consider the random reduced density matrix under the form $\rho_A=ZZ^{\dagger}/\Tr\left(ZZ^{\dagger}\right)$ \cite{nechita2007asymptotics}, where $Z$ is an $\alpha\times\beta$ random matrix for $1 \leq \alpha \leq \beta$. We have $E=-\sum^{\alpha-1}_{i=0}\lambda_{i}\ln{\lambda_{i}}$ where $\lambda_{i}$ are eigenvalues of $\rho_A$.

We can eliminate the trace and rewrite $\rho_A$ into a form that the MPD is applicable: $\rho_A=\frac{1}{\beta}XX^{\dagger}=\left(\frac{X}{\sqrt{\beta}}\right)\left(\frac{X}{\sqrt{\beta}}\right)^{\dagger}$, where $X$ is an $\alpha\times\beta$ random matrix with mean $0$ and $\sigma_0=\frac{1}{\sqrt{\alpha}}$. Therefore, $\frac{X}{\sqrt{\beta}}$ can be considered a "normalized" matrix, with coefficients $c_i$ that correspond to the quantum state $\ket{\psi}=\sum_{i=0}^{\alpha\beta-1}c_i\ket{i}$ with $\sum_{i=0}^{\alpha\beta-1}|c_i|^2=1$. We define $\lambda=\alpha/\beta$. The von Neumann entropy of a subsystem of dimension $\alpha$ can be written as

\begin{equation}
E=-\alpha\int x\ln{x}d\nu\left(x\right),
\label{eq_S_alpha_beta}
\end{equation}
with

\begin{equation}
d\nu\left(x\right)=\frac{1}{2\pi\sigma_0^2}\frac{\sqrt{\left(\lambda_{+}-x\right)\left(x-\lambda_{-}\right)}}{\lambda x}\mathbb{1}_{x\in[\lambda_-,\lambda_+]}dx,
\label{eq_dnu_ENT}
\end{equation}
and

\begin{equation}
\lambda_{\pm}=\sigma_0^2\left(1\pm \sqrt{\lambda}\right)^2.
\label{eq_lambda_ENT}
\end{equation}

Then we have

\begin{equation}
E=-\frac{\alpha}{2\pi\sigma_0^2\lambda}\int^{\lambda_+}_{\lambda_-}\ln{x}\sqrt{\left(\lambda_{+} -x\right)\left(x-\lambda_{-}\right)}dx,
\label{eq_E_alpha_beta_2}
\end{equation}
which can be simplified back into

\begin{equation}
E=\ln{\alpha}-\frac{\alpha}{2\beta},
\label{eq_S_alpha_beta_3}
\end{equation}
with the integral in Eq. (\ref{eq_A2})

\begin{widetext}
\begin{equation}
\begin{aligned}
&\int^b_a \ln{x} \sqrt{\left(b-x\right)\left(x-a\right)}dx\\
=&\frac{\pi}{16}\left(a^2+6ab+b^2-4\sqrt{ab}\left(a+b\right)-4\left(a-b\right)^2\ln{2}+2\left(a-b\right)^2\ln\left(a+b+2\sqrt{ab}\right)\right).
\end{aligned}
\label{eq_A2}
\end{equation}
\end{widetext}

\end{proof}

\subsection{Average entropy of a subsystem (general case)}\label{general}

In this section, we consider a slightly different variant when the mean of $X$ is not $0$ but $\gamma$. We still have $\lambda=\alpha/\beta$ and the eigenvalues of $\rho_A$ are still $\lambda_0, ..., \lambda_{\alpha-1}$ with $\lambda_0 \geq ... \geq \lambda_{\alpha-1}$. The standard deviation $\sigma_\gamma$ should be evaluated with $\gamma$ to maintain the sum of all eigenvalues as $1$. We have

\begin{equation}
\lambda_0=\alpha\gamma^2.
\end{equation}

The remaining eigenvalues $\lambda_1, ... ,\lambda_{\alpha-1}$ will follow the MPD, and their sum would be

\begin{equation}
\sum^{\alpha-1}_{i=1}\lambda_i\sim\alpha\int x d\nu(x)=\frac{\alpha}{2\pi\lambda\sigma_{\gamma}^2} \int_{\lambda_{-}}^{\lambda_{+}}\sqrt{\left(\lambda_{+}-x\right)\left(x-\lambda_{-}\right)}dx=\alpha\sigma_{\gamma}^2,
\end{equation}
with the integral in Eq. (\ref{eq_A1})

\begin{equation}
\int^b_a \sqrt{\left(b-x\right)\left(x-a\right)}dx=\frac{\pi}{8}\left(b-a\right)^2.
\label{eq_A1}
\end{equation}

To have the sum of all eigenvalues of $\rho_A$ equal $1$, we should have the relation

\begin{equation}
\alpha\left(\gamma^2+\sigma_{\gamma}^2\right)=1.
\label{eq_sum1}
\end{equation}

This relation can also be obtained using the assumption that the quantum state is normalized, or $\rho_A$ has the trace $1$. These three conditions are equivalent, and all originated from the fact that only pure quantum states are investigated in this article.

We can deduce that

\begin{equation}
\sigma_{\gamma}^2=\frac{1-\lambda_0}{\alpha},
\label{eq_variance}
\end{equation}
then use this expression to replace $\sigma_0$ in Eq. (\ref{eq_lambda_ENT}) and Eq. (\ref{eq_E_alpha_beta_2}). We then have the entropy depending on $\lambda_0$

\begin{equation}
E_{\lambda_0}=-\frac{\alpha}{2\pi\sigma_{\gamma}^2\lambda}\int^{\lambda_+}_{\lambda_-} \ln{x}\sqrt{\left(\lambda_{+} -x\right)\left(x-\lambda_{-}\right)}dx-\lambda_0\ln{\lambda_0},
\label{eq_E_alpha_beta_lambda_2}
\end{equation}
with

\begin{equation}
\lambda_{\pm}=\sigma_{\gamma}^2\left(1\pm \sqrt{\lambda}\right)^2.
\label{eq_lambda_ENT_2}
\end{equation}

After applying the integral in Eq. (\ref{eq_A2}) and massive reduction, we finally have

\begin{widetext}
\begin{equation}
E_{\lambda_0}=\left(1-\lambda_0\right)\left(\ln{\alpha}-\ln\left(1-\lambda_0\right)-\frac{\alpha}{2\beta}\right)-\lambda_0\ln{\lambda_0},
\label{eq_E_alpha_beta_lambda_3}
\end{equation}
\end{widetext}
which is the average entropy for a given $\lambda_0$. With this expression, we can recover Eq. (\ref{eq_E_alpha_beta_2}) for $\lambda_0\rightarrow 0$. When $\beta\rightarrow\infty$, we can recover the upper bound

\begin{equation}
\lim_{\beta\rightarrow\infty}E_{\lambda_0}\sim f_3.
\end{equation}

\begin{figure}
\centering
\includegraphics[width=0.48\linewidth]{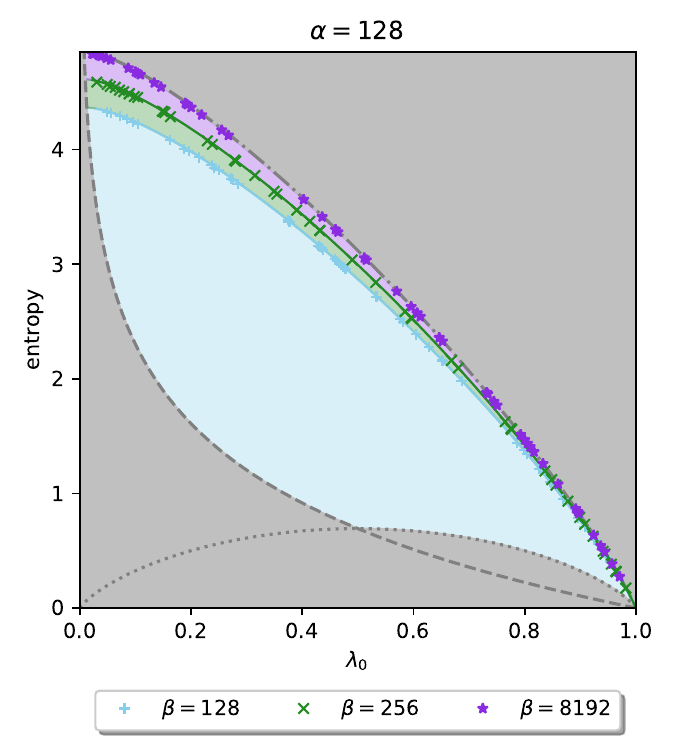}
\caption{The numerical results (in data points) versus the analytical solution (in line). The von Neumann entropy of three different $\beta\geq\alpha$ ($30$ data points for each $\beta$) for the same $\alpha=128$. Each of the $90$ data points in the plot is the entropy of a randomly generated normalized reduced density matrix under the form $\rho_A=ZZ^{\dagger}/\Tr\left(ZZ^{\dagger}\right)$.}
\label{alphabeta}
\end{figure}

In FIG. \ref{alphabeta} we compare our analytical solution with the numerical results.

\begin{figure}
\centering
\includegraphics[width=0.48\linewidth]{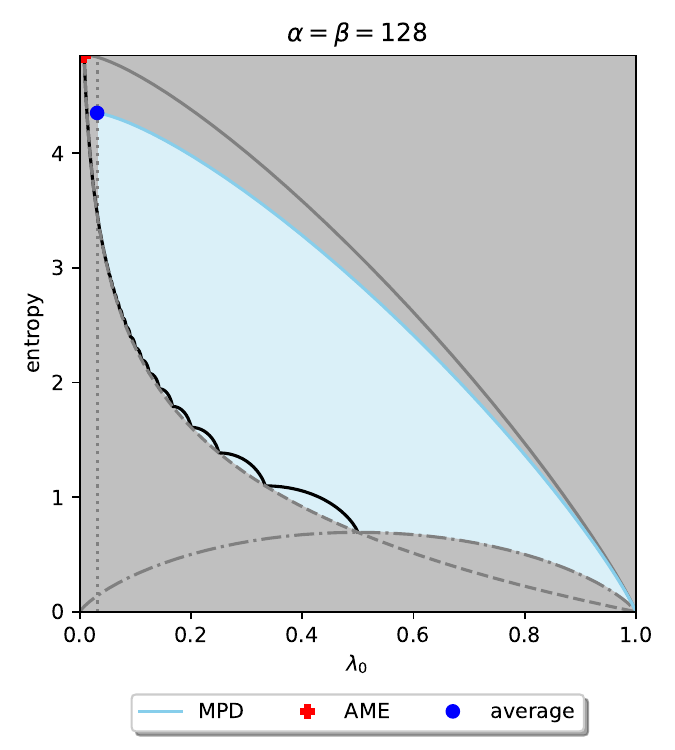}
\caption{The comparison between the tight and flexible boundaries for the entanglement trajectory defined with the von Neumann entropy. The vertical gray dotted line is $\lambda_+$ in Eq. (\ref{eq_lambda_ENT}) to indicate the lower bound of $\lambda_0$ for the flexible boundary, when the variance reaches the theoretical maximum, representing the case of average entropy.}
\label{AME}
\end{figure}

The result given by asymptotic analysis reduces the tight boundary to a more flexible boundary, as shown in FIG. \ref{AME}. In contrast to "tight," which implies a barrier forbidden by mathematics, "flexible" indicates a barrier protected by randomness. The absolutely maximally entangled (AME) states \cite{helwig2013absolutely,goyeneche2015absolutely} are rare and located at the extremity of the tight boundary if they exist \cite{AME_table}. And almost all quantum states (average entropy of a subsystem) are located at the extremity of the flexible boundary.

\subsection{Relation with previous examples}\label{relation}

\begin{figure}
\centering
\includegraphics[width=0.48\linewidth]{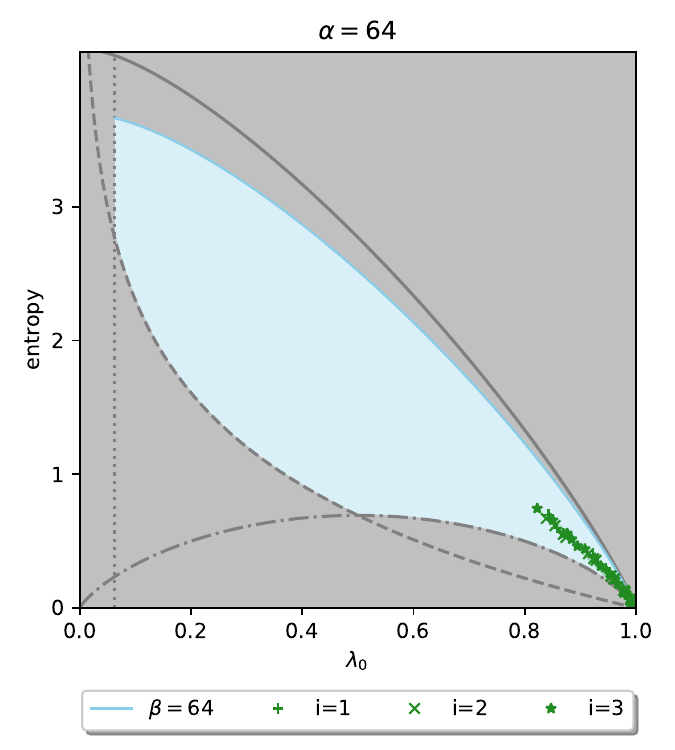}
\caption{The entanglement trajectories of adiabatic computation versus the boundaries.}
\label{EC}
\end{figure}

\begin{figure}
\centering
\includegraphics[width=0.48\linewidth]{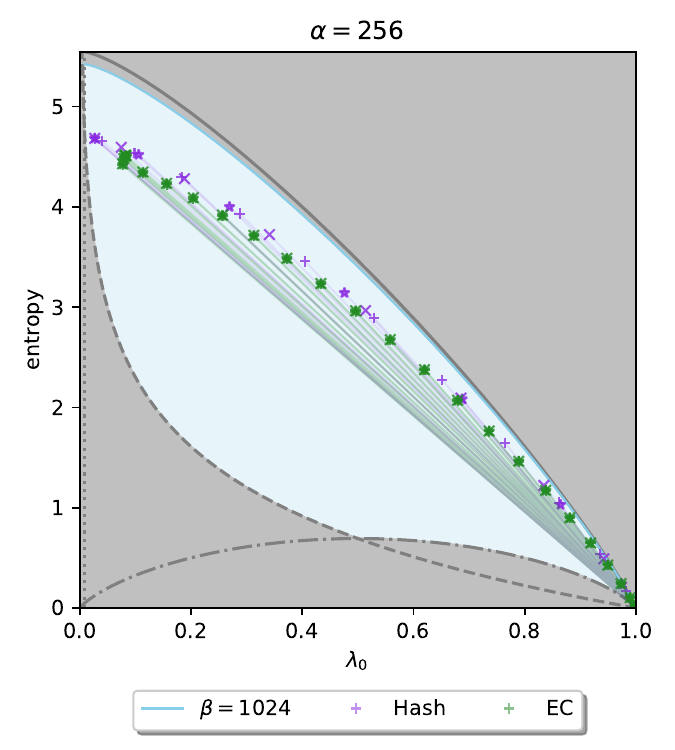}
\caption{The entanglement trajectories of the Grover algorithm versus the boundaries.}
\label{Grover}
\end{figure}

\begin{figure}
\centering
\includegraphics[width=0.48\linewidth]{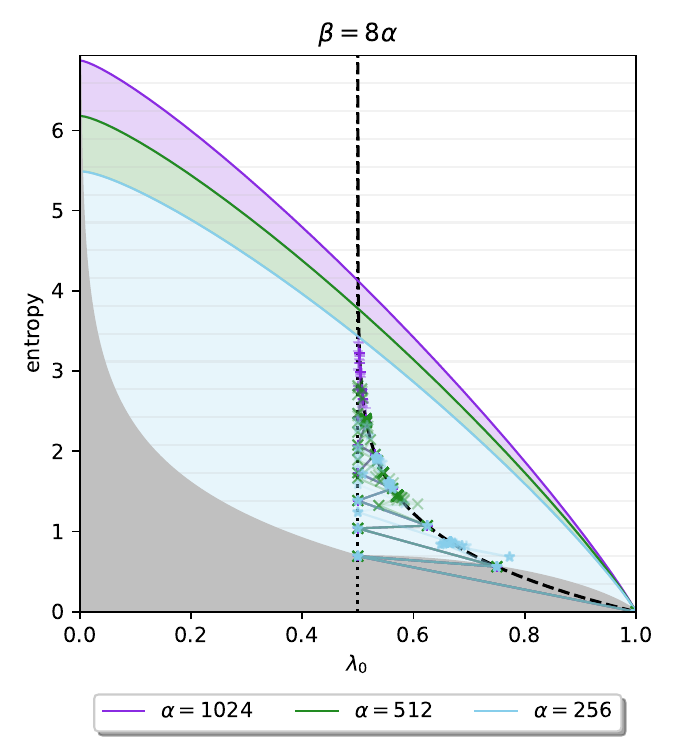}
\caption{The entanglement trajectories of the Shor algorithm versus the boundaries. Numerically the maximum entropy is becoming closer to the upper bound given by MPD as $n$ increases.}
\label{Shor}
\end{figure}

We re-plot FIG. \ref{EC_bis}, FIG. \ref{Grover_bis} and FIG. \ref{Shor_bis} with their flexible boundary, as shown in FIG. \ref{EC}, FIG. \ref{Grover} and FIG. \ref{Shor}. As we can see, the entanglement trajectories obtained from physical systems stay below the line deduced from MPD, which is a non-trivial result. Although it is difficult to theoretically or physically construct quantum states that are located above the line of average entropy over every possible bi-partition, such as AME states, preparing such a state for one bi-partition is easy. A possible explanation might be that reaching an entanglement higher than average will not help computation since the asymptotic analysis is derived from randomness. However, this argument is difficult to prove because counter-examples are not hard to formulate.

In particular, we have

\begin{equation}
E_{\lambda_0=\frac{1}{2}}=\frac{1}{2}\ln{\alpha}-\frac{\alpha}{4\beta}+\ln{2}.
\label{eq_l0.5}
\end{equation}

In the case of the Shor algorithm using $2n+3$ qubits. If we postulate that when the entanglement trajectory reaches the maximum as $\lambda_0=\frac{1}{2}$, and the entropy will be near and upper bounded by $E_{\lambda_0=\frac{1}{2}}$, then the maximum entropy that we expect from this particular partition of semi-classical Shor circuit would be $\frac{n+2}{2}\ln{2}-\frac{1}{32}$, by replacing $\alpha$ with $2^{n}$ and $\beta$ with $2^{n+3}$. That means the semi-classical Shor algorithm can not be simulated efficiently on a traditional computer. This example can be seen as an application of the asymptotic analysis.

\subsection{$k$-almost Prime states and their unions}

In the previous section \ref{relation}, we have demonstrated with examples that our asymptotic analysis defines a flexible upper bound for the entanglement trajectory. In this section, we provide another example to justify why this upper bound should not be lower. In this case, a series of quantum states are tracked to obtain the entanglement trajectory instead of those taken from different moments along a physical process.

In Ref. \cite{latorre2013quantum,latorre2014there,garcia2020prime}, the authors have studied the entanglement of the Prime state. The Prime state is defined as an equally weighted superposition of prime numbers in $n$ qubits computational basis

\begin{equation}
\ket{\mathbb{P}}=\frac{1}{\sqrt{\pi\left(2^n-1\right)}}\sum_{p \in\text{prime}}\ket{p},
\end{equation}
where each prime number $p=p_0 2^0 + p_1 2^1 + ... + p_{n-1} 2^{n-1}$ is implemented as $\ket{p}=\ket{p_{n-1},...,p_{0}}$, and $\pi\left(2^n-1\right)$ is the amount of prime numbers less than $2^n$. For example, when $n=4$, we have

\begin{equation}
\begin{aligned}
\ket{\mathbb{P}}&=\frac{1}{\sqrt{6}}\left(\ket{2}+\ket{3}+\ket{5}+\ket{7}+\ket{11}+\ket{13}\right)\\
                &=\frac{1}{\sqrt{6}}\left(\ket{00}\ket{10}+\ket{00}\ket{11}+\ket{01}\ket{01}+\ket{01}\ket{11}+\ket{10}\ket{11}+\ket{11}\ket{01}\right).
\end{aligned}
\end{equation}

The natural bi-partition entanglement entropy of the Prime state for even $n$ is investigated in Ref. \cite{latorre2014there} (the logarithm in base $2$ was used). In this example, we explore its other variant, referred to as the $k$-almost Prime state. $k$-almost Prime is a number that has $k$ prime factors, usually denoted as $p_k$. Here we define $k$-almost Prime state $\ket{\mathbb{P}_k}$ as the superposition

\begin{equation}
\ket{\mathbb{P}_k}=\frac{1}{\sqrt{\pi_k\left(2^n-1\right)}}\sum_{p_k\in\text{$k$-almost prime}}\ket{p_k},
\end{equation}
where $\pi_k\left(2^n-1\right)$ is the amount of $k$-almost Prime numbers less than $2^n$.

And their union is defined as the equal superposition of $\bigcup_{i=1}^k p_k$

\begin{equation}
\ket{\mathbb{U}_k}=\frac{1}{\sqrt{\pi_1\left(2^n-1\right)+...+\pi_k\left(2^n-1\right)}}\left(\sum_{p_1\in\text{$1$-almost prime}}\ket{p_1}+...+\sum_{p_k\in\text{$k$-almost prime}}\ket{p_k}\right).
\end{equation}

We have $\ket{\mathbb{P}}=\ket{\mathbb{P}_1}=\ket{\mathbb{U}_1}$. When $n=4$,

\begin{equation}
\ket{\mathbb{P}_2}=\frac{1}{\sqrt{6}}\left(\ket{4}+\ket{6}+\ket{9}+\ket{10}+\ket{14}+\ket{15}\right),
\end{equation}
and

\begin{equation}
\ket{\mathbb{P}_3}=\frac{1}{\sqrt{2}}\left(\ket{8}+\ket{12}\right),
\end{equation}
are $k$-almost Prime state. Also

\begin{equation}
\ket{\mathbb{U}_2}=\frac{1}{\sqrt{12}}\left(\ket{2}+\ket{3}+\ket{4}+\ket{5}+\ket{6}+\ket{7}+\ket{9}+\ket{10}+\ket{11}+\ket{13}+\ket{14}+\ket{15}\right),
\end{equation}
and

\begin{equation}
\ket{\mathbb{U}_3}=\frac{1}{\sqrt{14}}\left(\ket{2}+\ket{3}+\ket{4}+\ket{5}+\ket{6}+\ket{7}+\ket{8}+\ket{9}+\ket{10}+\ket{11}+\ket{12}+\ket{13}+\ket{14}+\ket{15}\right),
\end{equation}
are their union.

\begin{figure}
\centering
\includegraphics[width=0.48\linewidth]{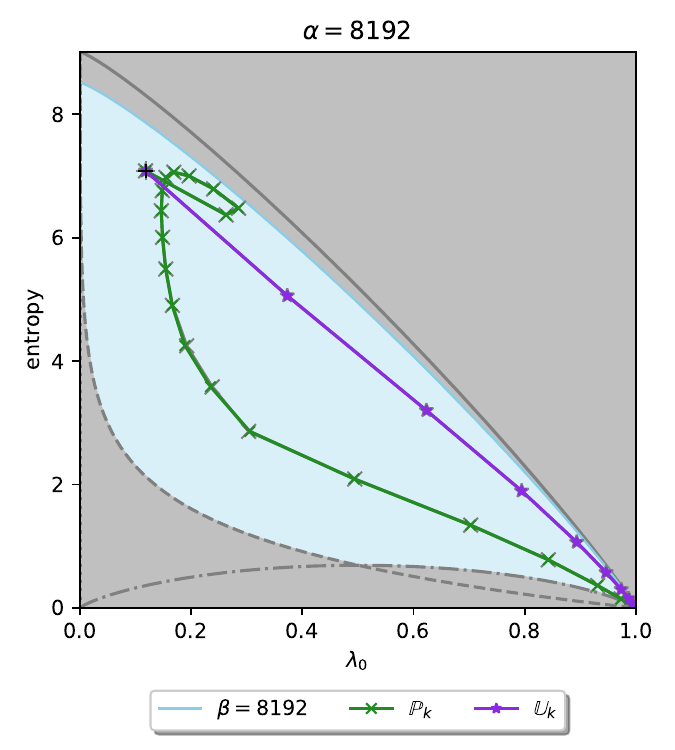}
\caption{The entanglement trajectories for $k$-almost Prime states and their unions when $n=26$. The Prime state $\ket{\mathbb{P}}$ is plotted with black $+$ and located on the intersection between $\ket{\mathbb{P}_k}$ and $\ket{\mathbb{U}_k}$. $QFT\ket{\mathbb{P}_k}$ and $QFT\ket{\mathbb{U}_k}$ are plotted with slightly bigger gray markers. They are very close to the data points of $\ket{\mathbb{P}_k}$ and $\ket{\mathbb{U}_k}$, therefore almost invisible.}
\label{Prime_Almost}
\end{figure}

In particular, we notice the fact that

\begin{equation}
\begin{aligned}
\ket{\mathbb{P}_{n-1}}&=\frac{1}{\sqrt{2}}\left(\ket{2^{n-1}}+\ket{3\times2^{n-2}}\right)\\
&=\frac{1}{\sqrt{2}}\left(\ket{1\underbrace{0...0}_{\frac{n}{2}-1}}\ket{ \underbrace{0...0}_{\frac{n}{2}}}+\ket{1 1\underbrace{0...0}_{\frac{n}{2}-2}}\ket{ \underbrace{0...0}_{\frac{n}{2}}}\right),
\end{aligned}
\end{equation}
of which the entanglement entropy is $0$. On the other hand, $\ket{\mathbb{U}_{n-1}}$ would be the equal superposition of all computational basis except $\ket{0}$ and $\ket{1}$, which also has $0$ entanglement entropy if we take the natural bi-partition. As a consequence, before we draw the entanglement trajectories of $\ket{\mathbb{P}_k}$ and $\ket{\mathbb{U}_k}$ for $k\in\{1,...,n-1\}$, we can already anticipate that they will join on the point of the Prime state as $k=1$, also on the point $\left(1,0\right)$ for $k=n-1$. The actual numerical simulation is shown in FIG. \ref{Prime_Almost} and supports our prediction. The entanglement trajectory is extremely close to our asymptotic analysis while staying strictly below it.

\section{General Entanglement Trajectory}

This article mainly focuses on the entanglement trajectory defined with the von Neumann entropy since it is the most widely used entropy in other research. However, different variants of entropy are only different functions defined on the eigenvalues of $\rho_A$. Their asymptotic analysis can be obtained with the same method in section \ref{general}, which allows us to define boundaries of different entanglement trajectories. Here we provide three examples.

\subsection{Entanglement trajectory in the momentum space}

If a quantum state $\ket{\psi}$ is given in position state, its Fourier transform $QFT\ket{\psi}$ gives the same state in the momentum space. Considering a random state $\ket{\psi}=\sum_{i=0}^{\alpha\beta-1}c_i\ket{i}$. Assuming that the dominant eigenvalues of its reduced density matrix $\rho_A$ is still $\lambda_0$, then the normalized state after performing the QFT on $\ket{\psi}$ can be written as

\begin{equation}
QFT\ket{\psi}=Q_0\ket{0}+\sum_{i=1}^{\alpha\beta-1}Q_i\ket{i},
\end{equation}
where $|Q_0|^2\sim\lambda_0$ and $Q_i$ are random variables with mean $0$ and variance $\sigma_{QFT}^2=\left(1-\lambda_0\right)/\alpha\beta$, which according to the Central limit theorem \cite{rosenblatt1956central}, can be approximated by $\mathcal{N}_{\mathbb{C}}\left(0,\sigma_{QFT}\right)$. The reduced density matrix of $QFT\ket{\psi}$ can be represented as

\begin{equation}
\rho_{QFT}=
\begin{pmatrix}
q_0 & q_2 & ... & q_2 & q_2\\
q_2 & q_1 & ... & q_3 & q_3\\
\vdots & \vdots & & \vdots & \vdots\\
q_2 & q_3 & ... & q_1 & q_3\\
q_2 & q_3 & ... & q_3 & q_1
\end{pmatrix},
\label{eq_Q}
\end{equation}
where

1) $q_0\sim\lambda_0$,

2) $q_1\sim\mathcal{N}\left(\left(1-\lambda_0\right)/\alpha,\sigma_1\right)$ with $\sigma_1\sim\mathcal{O}\left(\alpha\sigma_{QFT}^2\right)$,

3) $q_2\sim\mathcal{N}_\mathbb{C}\left(0,\sigma_2\right)$ with $\sigma_2\sim\sqrt{\lambda_0}\sigma_{QFT}$,

4) $q_3\sim\mathcal{N}_\mathbb{C}\left(0,\sigma_1\right)$.

\begin{figure}
\centering
\includegraphics[width=0.48\linewidth]{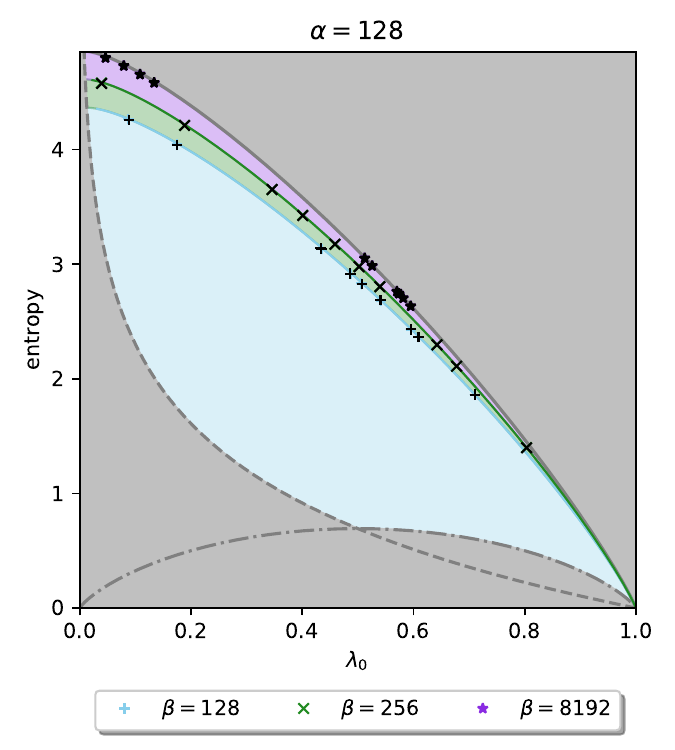}
\caption{The von Neumann entropy before and after applying the QFT of different $\beta\geq\alpha$ for the same $\alpha=128$, $10$ random states for each $\beta$. Notice that data points after applying the QFT are plotted with black markers and overlap completely with the original state.}
\label{QFT}
\end{figure}

The symbol $\mathcal{O}$ gives an order of approximation since the elements of $QFT\ket{\psi}$ are correlated and can not be treated as i.d.d random variables. When $\alpha\gg 0$, for a given $\lambda_0$, we have $q_0\gg q_1,q_2,q_3$, which indicate that $\lambda_0^{QFT}$ the dominant eigenvalues of $\rho_{QFT}$ is very likely to be near $\lambda_0$. We can observe from FIG. \ref{QFT} that $QFT\ket{\psi}$ and $\ket{\psi}$ share the exact dominant eigenvalue and entropy, which follows the analytical result in Eq. (\ref{eq_E_alpha_beta_lambda_3}). Therefore, the boundary of entanglement entropy defined with von Neumann entropy remains valid in the momentum space. Also, as we can observe from FIG. \ref{Prime_Almost}, the entanglement trajectory after performing QFT almost overlaps with the original one.

\subsection{Entanglement gap}

In this section, we investigate the relation between $\lambda_0$ and the entanglement gap. The reduced density matrix $\rho_A$ can be written as

\begin{equation}
\rho_A=e^{-\mathcal{H}_A},
\end{equation}
where $\mathcal{H}_A$ is called the entanglement Hamiltonian \cite{li2008entanglement,cirac2011entanglement,roy2021bulk,alba2021entanglement,calabrese2008entanglement,lauchli2010disentangling}. The entanglement spectrum levels $\xi_i=-\ln{\lambda_i}$ are the "energies" of $\mathcal{H}_A$. The entanglement gap is a natural quantity defined as

\begin{equation}
\delta\xi=\xi_1-\xi_0.
\end{equation}

With this definition and the approach that we deduced $f_1$, $f_2$ and $f_3$, we can have a tight boundary for entanglement gap confined by $g_1$, $g_2$ and $g_3$, such that

\begin{equation}
g_1=\ln{\lambda_0}-\ln\left(1-\lambda_0\right)
\end{equation}
is plotted with a gray dash-dotted line,

\begin{equation}
g_2=0,
\end{equation}
and

\begin{equation}
g_3=\ln{\lambda_0}-\ln\left(\frac{1-\lambda_0}{\alpha}\right)
\end{equation}
is plotted with a gray solid line.

\begin{figure}
\centering
\includegraphics[width=0.48\linewidth]{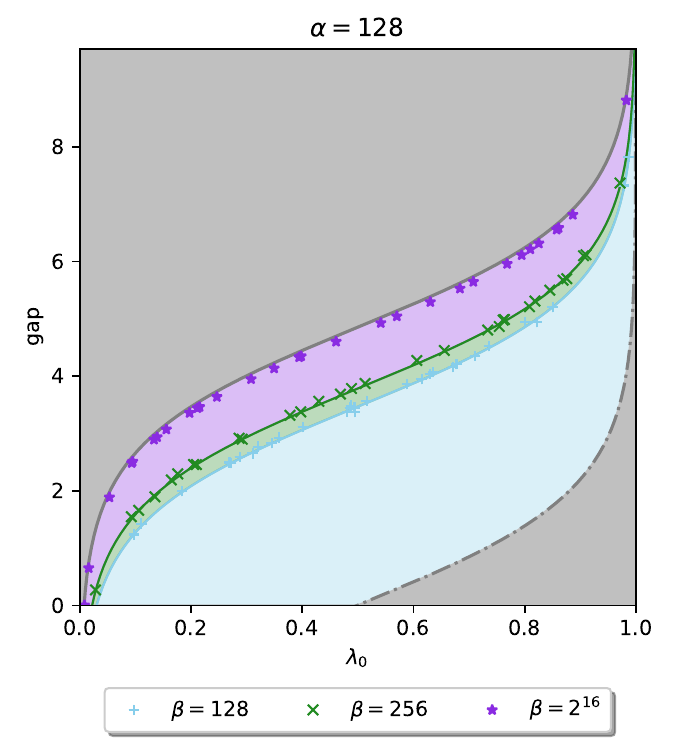}
\caption{The numerical simulation (in points) versus the analytical solution (in line). The entanglement gap for $90$ random reduced density matrices ($30$ per each $\beta$) and Eq. (\ref{eq_gap}).}
\label{Gap}
\end{figure}

To calculate the asymptotic value of $\delta\xi$, we take $\lambda_+$ the right end of the MPD as $\lambda_1$, which is $\sigma_\gamma^2\left(1+\sqrt{\lambda}\right)^2$, and $\delta\xi$ can be written as a function of $\alpha$, $\beta$ and $\lambda_0$ with Eq. (\ref{eq_variance})

\begin{equation}
\delta\xi=\ln{\lambda_0}-\ln\left(\sigma_\gamma^2\left(1+\sqrt{\lambda}\right)^2\right)=\ln{\lambda_0}-\ln\left(\frac{1-\lambda_0}{\alpha}\left(1+\sqrt{\frac{\alpha}{\beta}}\right)^2\right),
\label{eq_gap}
\end{equation}
and shown in FIG. \ref{Gap}. When $\beta\rightarrow\infty$, we have

\begin{equation}
\lim_{\beta\rightarrow\infty}\delta\xi=g_3.
\end{equation}

\begin{figure}
\centering
\includegraphics[width=0.48\linewidth]{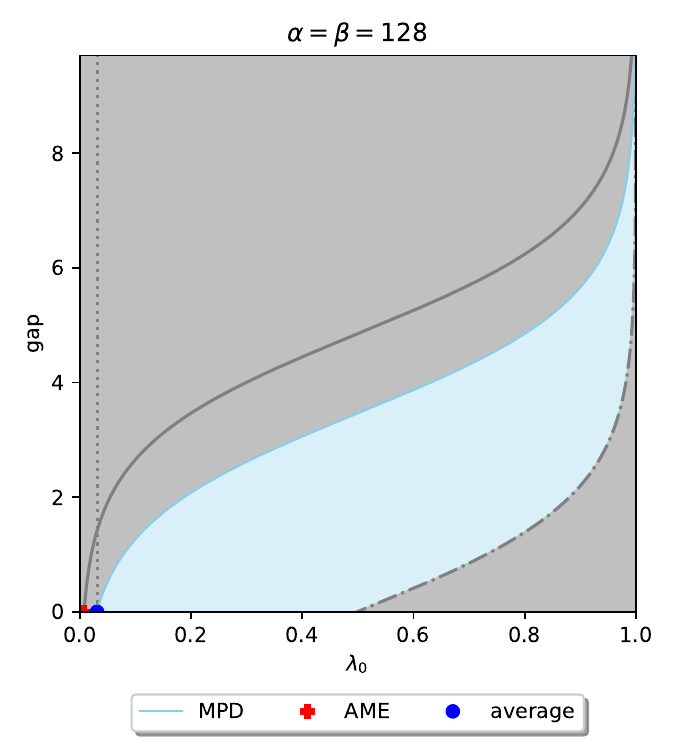}
\caption{The comparison between the tight and flexible boundaries entanglement trajectory defined with the entanglement gap.}
\label{AME_gap}
\end{figure}

From FIG. \ref{alphabeta} and FIG. \ref{Gap}, we can observe a monotone decreasing relation between $\lambda_0$ and $E_{\lambda_0}$, and a monotone increasing relation between $\lambda_0$ and $\delta\xi$. These two relations imply that the von Neumann entropy decreases with the entanglement gap, as stated by Li and Haldane in 2008 with the ideal Moore-Read state \cite{li2008entanglement}. Furthermore, the trajectory boundaries defined with von Neumann entropy (FIG. \ref{AME}) and the entanglement gap share a similarity. Intuitively, we can obtain FIG. \ref{AME_gap} by pulling down the left side of FIG. \ref{AME} to $0$ and pushing up the right side to $\infty$.

\begin{figure}[t]
\begin{subfigure}[t]{0.48\linewidth}
\centering
\includegraphics[width=0.98\linewidth]{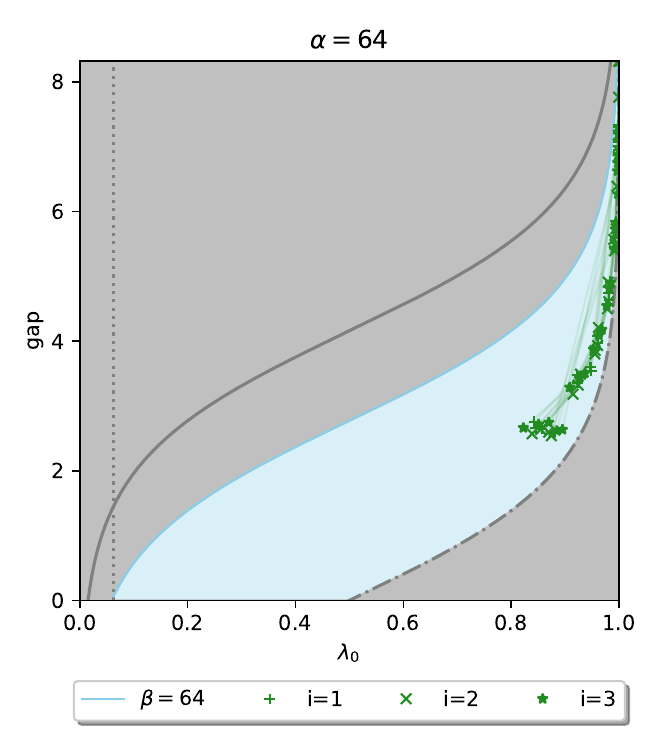}
\caption{Corresponding to FIG. \ref{EC}, adiabatic quantum computation to solve EC problem.}
\label{EC_gap}
\end{subfigure}
\hfill
\begin{subfigure}[t]{0.48\linewidth}
\includegraphics[width=1\linewidth]{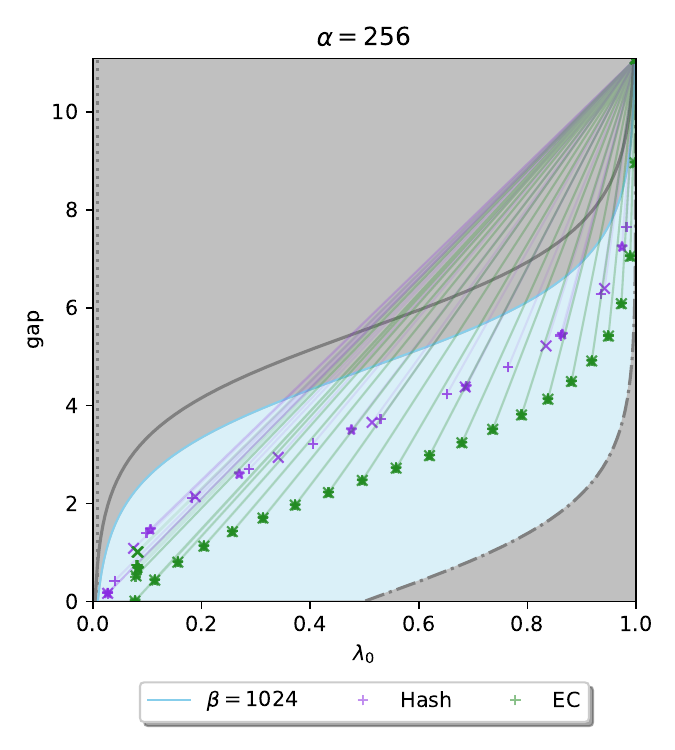}
\caption{Corresponding to FIG. \ref{Grover}, Grover algorithm to solve EC problem and the Hash function.}
\label{Grover_gap}
\end{subfigure}
\\
\begin{subfigure}[t]{0.48\linewidth}
\includegraphics[width=1\linewidth]{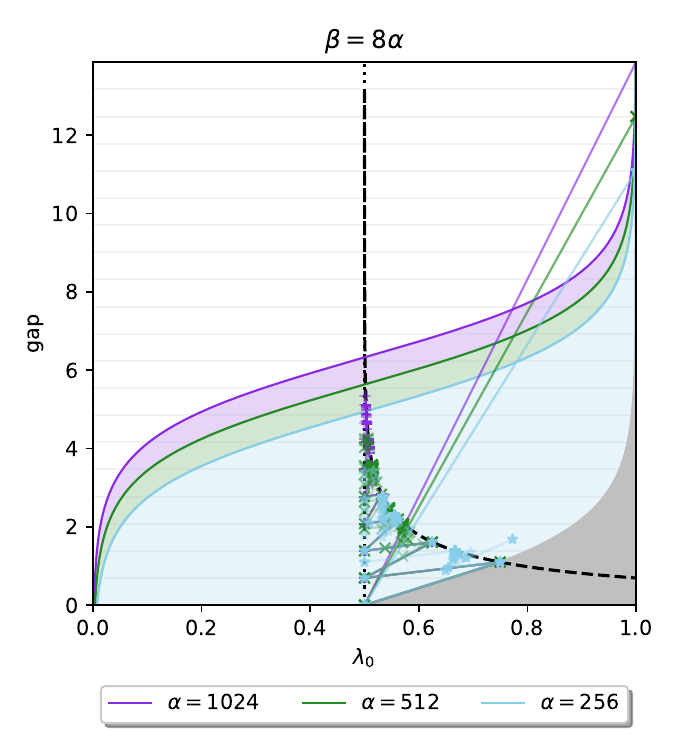}
\caption{Corresponding to FIG. \ref{Shor}, Shor algorithm. Gray horizontal lines are plotted with interval $\ln{2}$.}
\label{Shor_gap}
\end{subfigure}
\hfill
\begin{subfigure}[t]{0.48\linewidth}
\includegraphics[width=1\linewidth]{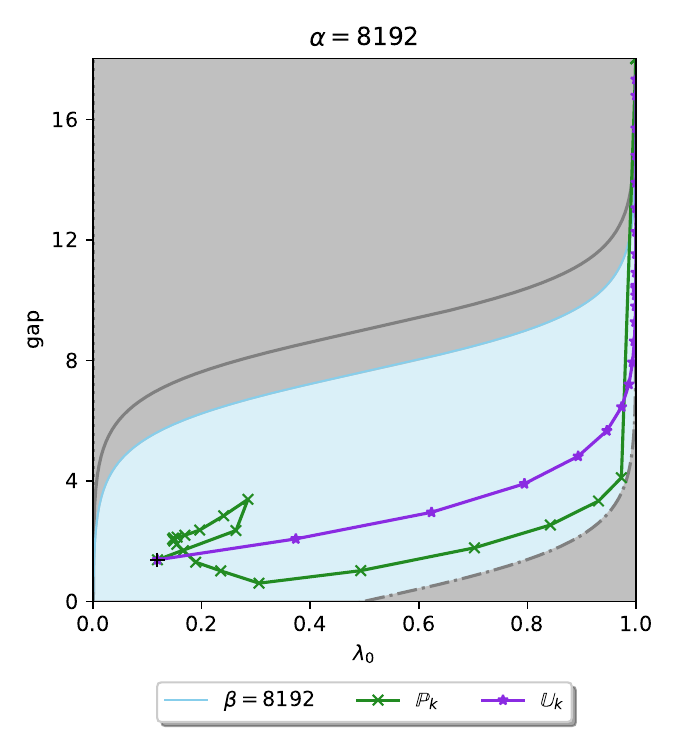}
\caption{Corresponding to FIG. \ref{Prime_Almost}, $k$-almost Prime states and their unions.}
\label{Prime_Almost_gap}
\end{subfigure}
\caption{Entanglement trajectory of previous examples with the entanglement gap as $y$-axis. For simplicity the point $\left(1,\infty\right)$ is replaced by $\left(1,2\ln{\alpha}\right)$.}
\label{Gap_examples}
\end{figure}

The trajectories of previous examples are re-plotted with the entanglement gap in FIG. \ref{Gap_examples}. In the example of semi-classical Shor algorithm, $f_{shor}$ is reformulated into

\begin{equation}
g_{shor}=\ln{x}+\ln{2}-\ln\left(2x-1\right).
\end{equation}

Also, Numerically, the maximum entanglement gap is becoming closer to the upper bound given by MPD as $n$ increases.

Three non-trivial observations can be obtained from this comparison. Firstly, the trajectories still remain within the flexible boundary even though, mathematically speaking, they are allowed to step outside. Secondly, since there is a monotone decreasing relation with $E_{\lambda_0}$ and $\delta\xi$ of the boundary, a smaller entanglement gap is expected with higher entropy. The simulation results do not fit this prediction. Instead, the figure of merit is the relative distance to the boundary. Data points that are closer to the boundary when the trajectory is defined with the von Neumann entropy remain closer to the corresponding boundary when the trajectory is defined with the entanglement gap. Furthermore, comparing FIG. \ref{Grover}, FIG. \ref{Shor} and (b), (c) of FIG. \ref{Gap_examples}, trajectories closer to the boundary defined by MPD present more value for quantum computation. Thirdly, the trajectory preserves its particular "topology" in each example over different entanglement measures. Intuitively, we pull the trajectory down on the left side and push up on the right. These three conclusions are distinct, but they support each other.

Here we highlight that these three properties are observations which emerge naturally from simulations of quantum computing and are impossible to prove mathematically. A trajectory plotted with von Neumann entropy does not correspond to a unique path when plotted with the entanglement gap, and vice versa. Counterexamples of these properties can always be designed. In the abstract formulation, the eigenvalues $\left(\lambda_0,...,\lambda_{\alpha-1}\right)$ locate on a compact $\alpha-1$ dimension surface in the space $\mathbb{R}^{\alpha}$ constrained by $\lambda_0 \geq...\geq\lambda_{\alpha-1} \geq 0$ and $\sum_{i=0}^{\alpha-1} \lambda_i=1$. The von Neumann entropy and the entanglement gap are projections of this high dimensional surface on $\mathbb{R}_+$, and most information about $\rho_A$ is lost. The unquoted word topology should not be used without a proper definition of homomorphism or homotopy.

These observations grant the entanglement trajectory an intrinsic physical value. Instead of just a tool for visualization, it reveals invariant properties over several examples with a variant measurement of entanglement, which is not guaranteed by mathematics.

\subsection{Different degrees of R\'{e}nyi entropy}

In this section, we study the entanglement trajectory defined with different degrees of R\'{e}nyi entropy. To simplify the analysis, we keep $\alpha=\beta$.

The R\'{e}nyi entropy is defined as \cite{nielsen2002quantum}

\begin{equation}
\begin{aligned}
E_d: \mathbb{R}_+^{\alpha} \times \mathbb{R}_+ &\rightarrow \mathbb{R}_+\\
\left(\lambda_0,...,\lambda_{\alpha-1}\right)\times d &\mapsto\frac{1}{1-d}\ln\left(\sum^{\alpha-1}_{i=0}\lambda_i^d\right),
\end{aligned}
\end{equation}
where $d\neq 1$ is the degree, and $\lambda_i$ are the eigenvalues of $\rho_A$, with $\lambda_0 \geq ... \geq \lambda_{\alpha-1}$ and $\sum_{i=0}^{\alpha-1}\lambda_i = 1$. When $d=0$, $E_d=\ln{\alpha}$ and called Hartley entropy or max-entropy. Notice $E_d$ behaves very differently from $d\rightarrow 0$ to $d=0$. When $d \rightarrow 1$, $E_d$ is the von Neumann entropy $-\sum_{i=0}^{\alpha-1}\lambda_i\ln{\lambda_i}$ \cite{nielsen2011r}, when $d = 2$, $E_d$ is the R\'{e}nyi entropy or the Collision entropy. Furthermore, when $d \rightarrow \infty$, $E_d$ is the min-entropy $-\ln{\lambda_0}$, which is another reason why $\lambda_0$ should be the $x$-axis, that it naturally gives the asymptotic analysis of R\'{e}nyi entropy before we define entanglement trajectory.

Since the series of R\'{e}nyi entropy are continuous extensions of the von Neumann entropy of different $d$, we will no elaborate on the corresponding $f_1$, $f_2$ and $f_3$ in detail and focus on the asymptotic value, which for $d\rightarrow1$ and $d=2$ can be calculated analytically by integrating over the MPD using Eq. (\ref{eq_A4})

\begin{widetext}

\begin{equation}
\int^b_0 \ln{x}\sqrt{x\left(b-x\right)}dx=\frac{\pi}{16}\left(2b^2\ln{b}-b^2\left(4\ln{2}-1\right)\right).
\label{eq_A4}
\end{equation}

We have

\begin{equation}
\begin{aligned}
\lim_{d\rightarrow1}E_d&=-\frac{\alpha}{2\pi\sigma_\gamma^2} \int_0^{4\sigma_\gamma^2}\ln{x}\sqrt{\left(4\sigma_\gamma^2 -x\right)x}dx-\lambda_0\ln{\lambda_0}\\
&=\left(1-\lambda_0\right)\left(\ln{\alpha}-\ln\left(1-\lambda_0\right)-\frac{1}{2}\right)-\lambda_0\ln{\lambda_0},
\end{aligned}
\label{eq_renyi_1}
\end{equation}
\end{widetext}
and

\begin{equation}
\begin{aligned}
E_2&=-\ln\left(\frac{\alpha}{2\pi\sigma_\gamma^2}\int_0^{4\sigma_\gamma^2}x\sqrt{\left(4\sigma_\gamma^2 -x\right)x}dx+\lambda_0^2\right)=-\ln\left(\frac{\lambda_0^2(\alpha+2)-4\lambda_0+2}{\alpha}\right).
\end{aligned}
\end{equation}

\begin{figure}[h]
\centering
\includegraphics[width=0.48\linewidth]{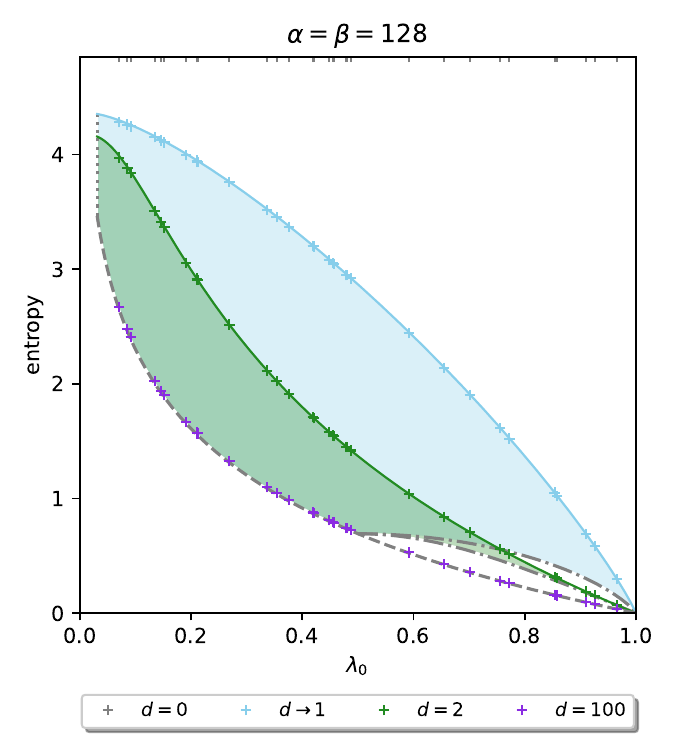}
\caption{The numerical simulation (in $+$) versus the analytical solution (in line). The R\'{e}nyi entropy with different $d$ (here we take $d=1.001$ for $d\rightarrow 1$) for the same $30$ random reduced density matrices and the analytic results.}
\label{Renyi}
\end{figure}

These three equations are verified with numerical results in FIG. \ref{Renyi}. We can recover Eq. (\ref{eq_renyi_1}) from Eq. (\ref{eq_E_alpha_beta_lambda_3}) by setting $\alpha=\beta$. R\'{e}nyi entropy with higher integer degrees can be obtained using integrals in TABLE. \ref{theory}.

\begin{table}
\centering
\begin{tabular}{ |c|c|c|c|c|c| } 
\hline
$d$ & 2 & 3 & 4 & 5 & 6\\ 
\hline
$\int_0^{t}x^{d-1}\sqrt{\left(t-x\right)x}dx$ & $\frac{1}{16}\pi t^3$ & $\frac{5}{128}\pi t^4$ & $\frac{7}{256}\pi t^5$ 
& $\frac{21}{1024}\pi t^6$ & $\frac{33}{2048}\pi t^7$\\
\hline
\end{tabular}
\caption{Integration table of $\int_0^{t}x^{d-1}\sqrt{\left(t-x\right)x}dx$.}
\label{theory}
\end{table}

Furthermore, each set of eigenvalues can uniquely define a path parameterized by $d$

\begin{equation}
\begin{aligned}
\mathbb{R}_+^{\alpha}  &\rightarrow \mathbb{R}_+^3\\
\left(\lambda_0,...,\lambda_{\alpha-1}\right) &\mapsto \left(\lambda_0,E_d,d\right).
\end{aligned}
\end{equation}

\begin{figure}
\centering
\includegraphics[width=1\linewidth]{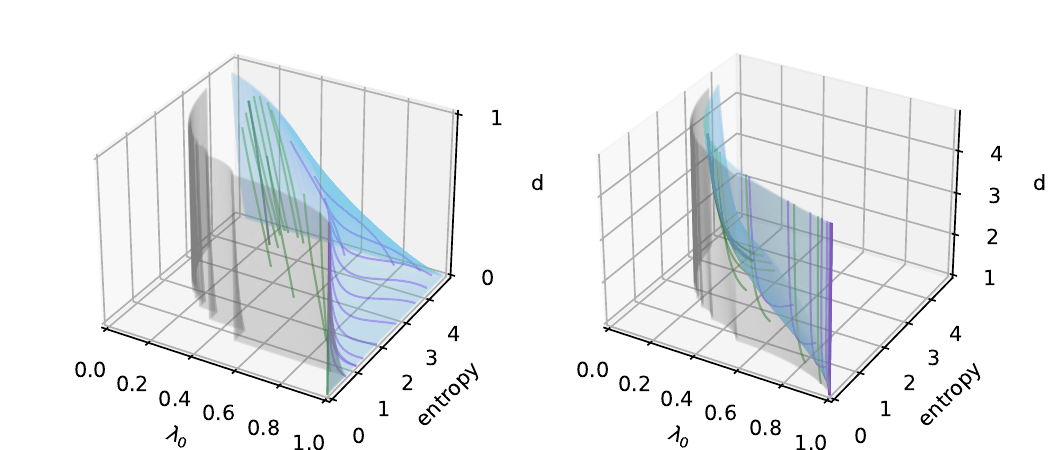}
\caption{The 3D entanglement trajectory of $k$-almost Prime states and their unions for $n=14$. $\mathbb{P}_k$ is plotted in green and $\mathbb{U}_k$ is plotted in violet. We have $\alpha=\beta=128$. The trajectory stays within the boundary, which is given by numerical approximation. In particular, the analytical boundary for $d \rightarrow 1$ and $d=2$ is defined in FIG. \ref{Renyi}.}
\label{3D}
\end{figure}

Therefore, instead of a data point on a 2D figure, a quantum state that is partially traced over a subsystem can be represented by a continuous and differentiable path located on the plane sectioned by $x=\lambda_0$. As shown in FIG. \ref{3D}, we can have a 3D version of the entanglement trajectory defined by R\'{e}nyi entropy, which is a extension of the 2D version and preserve its properties. We remind our readers that in the 2D version, lines that connect data points do not have a physical meaning. For the same reason, the surface formed by paths in this 3D version does not process any physical meaning either.

\section{Conclusion}

We introduce a new way, the entanglement trajectory, to visualize the transition of entanglement along a quantum process or a series of quantum states. We set a tight boundary to this trajectory from the definition of entropy and reduce it to a flexible boundary with asymptotic analysis based on random matrix theory. This part of the research remains valid for fundamental quantum physics research outside the context of quantum computation. Then, for several well-known examples from quantum computation, we simulate their entanglement trajectory and compare them with analytical boundaries. We observe some invariant properties, such that each trajectory preserves its "topology" and stays within the flexible boundary over variant scenarios and measurements of entanglement. These properties cannot be derived directly from the mathematical definition of entropy. Hence instead of a simple geometric visualization, the entanglement trajectory processes an underlying physical meaning and can provide more information about the quantum system than only studying the entanglement. It has the potential to help us understand more about the advantages of quantum computation since interesting instances reach near the boundary. Furthermore, the study of entanglement trajectory is inevitable if the advantage is truly justifiable because it provides a more general description of entanglement, the most distinguishable feature between classic and quantum despite different computation models.

\FloatBarrier

\begin{acknowledgments}
The author would like to thank Prof. J. I. Latorre and Prof. G. Sierra for their helpful information and discussion. Also, the author would like to thank S. Ramos-Calderer, who generously shared his code on an open-access platform. Finally, the author is grateful for a mathematical hint from one anonymous reviewer.
\end{acknowledgments}

\bibliographystyle{plain}

\end{document}